\documentclass[twocolumn,showpacs,preprintnumbers,amsmath,amssymb,APSl,prd,nofootinbib,superscriptaddress]{revtex4}
\usepackage{graphicx,color}
\usepackage{amsmath}
\usepackage{amssymb}
\usepackage{hyperref}
\usepackage[utf8]{inputenc}
\usepackage[english]{babel}
\usepackage{epsfig}
\usepackage{subfigure}
\usepackage{wasysym}
\usepackage{color,xcolor}
\usepackage{amsmath}
\usepackage{bm}
\usepackage{epsfig}
\usepackage{amsfonts}
\usepackage{dcolumn}
\usepackage{float}
\hypersetup{colorlinks=true, linkcolor=blue, citecolor=green}

\usepackage{dcolumn}% Align table columns on decimal point
\usepackage{bm}% bold math
\usepackage{ifpdf}
\usepackage{hyperref}
\usepackage{float}
\usepackage{bm}
\usepackage{xcolor,color,graphicx,graphics}
\usepackage[OT1]{fontenc}
\usepackage{latexsym,amssymb,amsmath,amsfonts}
\usepackage{makeidx}
\usepackage{epsfig}
\usepackage{epstopdf}
\usepackage{mathrsfs}
\hypersetup{colorlinks=true, linkcolor=blue, citecolor=green}
\usepackage{enumerate}
\usepackage{xcolor}
 \usepackage{multirow}
\begin{document}

%%%%%%%%%%%%%%%%%%%%%%%%%%%%%%%%%%%%%%%%%%%%%%%%%%%
%\title{Flow of matter near a black hole with spontaneous Lorentz symmetry breaking\\Matter flow near black holes with spontaneous Lorentz symmetry breaking
\title{Accretion dynamics in black holes with spontaneous Lorentz symmetry breaking}
%%%%%%%%%%%%%%%%%%%%%%%%%%%%%%%%%%%%%%%%%%%%%%%%%%%

%%%%%%%%%%%%%%%%%%%%%%%%%%%%%%%%%%%%%%%%%%%%%%%%%%%
\author{Daniela S. J. Cordeiro} 
        \email{fc52853@alunos.ciencias.ulisboa.pt}        
\affiliation{Instituto de Astrof\'{i}sica e Ci\^{e}ncias do Espa\c{c}o, Faculdade de Ci\^{e}ncias da Universidade de Lisboa, Edifício C8, Campo Grande, P-1749-016 Lisbon, Portugal}

	\author{Ednaldo L. B. Junior} \email{ednaldobarrosjr@gmail.com}
\affiliation{Faculdade de F\'{i}sica, Universidade Federal do Pará, Campus Universitário de Tucuruí, CEP: 68464-000, Tucuruí, Pará, Brazil}
\affiliation{Programa de P\'{o}s-Gradua\c{c}\~{a}o em F\'{i}sica, Universidade Federal do Sul e Suldeste do Par\'{a}, 68500-000, Marab\'{a}, Par\'{a}, Brazill}

     \author{José Tarciso S. S. Junior}
    \email{tarcisojunior17@gmail.com}
\affiliation{Faculdade de F\'{i}sica, Programa de P\'{o}s-Gradua\c{c}\~{a}o em F\'{i}sica, Universidade Federal do Par\'{a}, 66075-110, Bel\'{e}m, Par\'{a}, Brazill}

	\author{Francisco S. N. Lobo} \email{fslobo@ciencias.ulisboa.pt}
\affiliation{Instituto de Astrof\'{i}sica e Ci\^{e}ncias do Espa\c{c}o, Faculdade de Ci\^{e}ncias da Universidade de Lisboa, Edifício C8, Campo Grande, P-1749-016 Lisbon, Portugal}
\affiliation{Departamento de F\'{i}sica, Faculdade de Ci\^{e}ncias da Universidade de Lisboa, Edif\'{i}cio C8, Campo Grande, P-1749-016 Lisbon, Portugal}

    \author{Jorde A. A. Ramos} \email{jordealves@ufpa.br}
\affiliation{Faculdade de F\'{i}sica, Programa de P\'{o}s-Gradua\c{c}\~{a}o em F\'{i}sica, Universidade Federal do Par\'{a}, 66075-110, Bel\'{e}m, Par\'{a}, Brazill}

    \author{Manuel E. Rodrigues} \email{esialg@gmail.com}
\affiliation{Faculdade de F\'{i}sica, Programa de P\'{o}s-Gradua\c{c}\~{a}o em F\'{i}sica, Universidade Federal do Par\'{a}, 66075-110, Bel\'{e}m, Par\'{a}, Brazill}
\affiliation{Faculdade de Ci\^{e}ncias Exatas e Tecnologia, Universidade Federal do Par\'{a}, Campus Universit\'{a}rio de Abaetetuba, 68440-000, Abaetetuba, Par\'{a}, Brazil}

 \author{Diego Rubiera-Garcia} \email{ drubiera@ucm.es}
\affiliation{Departamento de Física Téorica and IPARCOS, Universidad Complutense de Madrid, E-28040 Madrid, Spain}

     \author{Luís F. Dias da Silva} 
        \email{fc53497@alunos.fc.ul.pt}
\affiliation{Instituto de Astrof\'{i}sica e Ci\^{e}ncias do Espa\c{c}o, Faculdade de Ci\^{e}ncias da Universidade de Lisboa, Edifício C8, Campo Grande, P-1749-016 Lisbon, Portugal}

    \author{Henrique A. Vieira} \email{henriquefisica2017@gmail.com}
\affiliation{Faculdade de F\'{i}sica, Programa de P\'{o}s-Gradua\c{c}\~{a}o em F\'{i}sica, Universidade Federal do Par\'{a}, 66075-110, Bel\'{e}m, Par\'{a}, Brazill}
%%%%%%%%%%%%%%%%%%%%%%%%%%%%%%%%%%%%%%%%%%%%%%%%%%%

%%%%%%%%%%%%%%%%%%%%%%%%%%%%%%%%%%%%%%%%%%%%%%%%%%%
\begin{abstract}
We investigate the spherical accretion of various types of fluids onto a Schwarzschild-like black hole solution modified by a Kalb-Ramond field implementing spontaneous Lorentz symmetry violation (LV). The system is analyzed for isothermal fluids characterized by the equation of state $p=\omega\rho$, including ultra-stiff, ultra-relativistic, and radiation fluids. We investigate the effect of the LV parameter $l$ on the fluid density $\rho(r)$, radial velocity $u(r)$, and accretion rate $\dot{M}$. Using a Hamiltonian dynamical systems approach, we examine the behavior near critical points and identify the sonic transitions in each scenario. Our results show that the LV parameter influences the location of critical points, the flow structure, and the accretion rate, with $l>0$ ($l<0$)  enhancing (suppressing) the latter. For ultra-stiff fluids, no critical points are found, and the flow remains entirely subsonic. For ultra-relativistic and radiation fluids, transonic solutions exist, with the position of the sonic point depending on the sign of $l$. We also analyze polytropic fluids $p=\mathcal{K}
\rho^{\Gamma}$ with $\Gamma=5/3$ and $\Gamma=4/3$, observing similar qualitative behavior, where the sonic transition is affected by both the equation of state and the LV parameter. These findings suggest that Lorentz symmetry breaking can significantly alter accretion dynamics in black hole spacetimes.
\end{abstract}
%%%%%%%%%%%%%%%%%%%%%%%%%%%%%%%%%%%%%%%%%%%%%%%%%%%

%\pacs{04.70.BW, 04.70.-s}
\date{\today}

\maketitle

%%%%%%%%%%%%%%%%%%%%%%%%%%%%%%%%%%%%%%%%%%%%%%%%%%%
\section{Introduction}
%%%%%%%%%%%%%%%%%%%%%%%%%%%%%%%%%%%%%%%%%%%%%%%%%%%

Local phenomena play a fundamental role in General Relativity (GR), for instance, in the  accretion of matter onto black holes \cite{Thorne:1974ve}. Any form of matter or radiation lurking close enough to a black hole is inevitably drawn into the event horizon, leading to an increase in the black hole’s mass. This process occurs widely throughout the universe and its description serves as an important tool for testing gravitational theories. The study of matter accretion was initiated by Bondi \cite{Bondi} within the framework of Newtonian mechanics, where solutions were obtained showing a transition of the fluid from subsonic to supersonic speeds. Subsequent investigations extended this analysis to spherically symmetric spacetimes in a cosmological context \cite{Bondi2, Malec, Malec2}. Relativistic effects were later incorporated by Michel in \cite{Michel}, giving rise to what is now known as Michel-like accretion, whose stability was discussed in \cite{Moncrief}. A significant contribution was made by Babichev and collaborators \cite{Babichev, Babichev2}, who demonstrated that if the accreting fluid is of the phantom type - an exotic form of energy with negative pressure - the black hole may lose mass instead of gaining it, potentially leading to its complete evaporation in a universe approaching a Big Rip scenario. Furthermore, it was shown in \cite{Jamil} that primordial black holes evaporate more rapidly when accreting phantom energy, highlighting the crucial role of the equation of state in accretion dynamics. These findings raise the hypothesis that supermassive black holes residing at the centres of giant galaxies could have formed through prolonged accretion of matter over cosmic timescales.

The process of fluid accretion onto black holes has been extensively studied using the framework of dynamical systems, both inside GR and beyond of it due to the potential imprints that new gravitational dynamics might leave on it. Indeed, it accretion dynamics has been studied within the context of modified theories of gravity such as $f(T)$ \cite{Ahmed, ednaldo1}, $f(R)$ \cite{Pun:2008ae,Amed2}, Einstein-aether \cite{Mukherjee:2024hht}, metric-affine gravity \cite{Mustafa:2023ngp}, massive gravity \cite{Panotopoulos:2021ezt}, scalar-tensor gravity \cite{John:2019was}, or Einstein-dilaton-Gauss-Bonnet gravity \cite{Maselli:2014fca,Zhang:2017unx}, and can also be employed to distinguish black holes from alternative horizonless compact objects \cite{Olivares:2018abq}. In this approach, the evolution of the system is analyzed through a Hamiltonian formulation, which enables the identification of sonic points, that is, critical locations where the radial velocity of the fluid equals the local speed of sound. These sonic points play a central role in determining the structure and classification of accretion flows. These studies have shown that the behavior of the fluid can be systematically categorized based on the equation of state of the isothermal fluid and the specific characteristics of the underlying black hole spacetime. This methodology offers a robust and transparent understanding of how accretion takes place in diverse gravitational settings. Within this issue, a result worth mentioning is the one of \cite{Abdul}, where the accretion dynamics was investigated for Kehagias-Sfetsos black holes as well as regular black holes described by a Dagum distribution function. It was found that the existence and location of sonic points, along with the qualitative features of the fluid flow, depend sensitively on the equation of state parameter, in agreement with theoretical predictions.

For the sake of this work we are interested on those gravitational models incorporating spontaneous Lorentz symmetry violation (LV) (see \cite{Addazi:2021xuf} for an overview of these theories). One possible mechanism involves the Kalb-Ramond (KR) tensor field acquiring a non-zero vacuum expectation value through a non-minimal coupling with the Ricci scalar, thereby inducing spontaneous symmetry breaking. The KR field has been extensively studied in a variety of physical contexts \cite{KRoriginal, VLorentz, VLorentz2, VLorentz3, KRteste, KRinfla, KRlenteforte, KRKumar, KRparticulas, KRparity, KRdarkmatter, KReletrico, KRtermo, KRacoplado1, KRMaluf}, and continues to attract increasing attention. Recently, in \cite{nosso2}, the effect of gravitational lensing was analyzed for a Schwarzschild-type black hole spacetime modified by the presence of a KR background field via a new single extra parameter $l$, as introduced in \cite{KR}. It was shown that an increase in the spontaneous symmetry breaking parameter leads to a reduction in the deflection angle of light. Based on observational data from Sgr A*, key astrophysical quantities such as image position, luminosity, and the Shapiro time delay were computed. Moreover, in \cite{nosso3}, time-like geodesics around spherically symmetric black holes within the KR-modified gravity framework were investigated. It was found that the parameter $l$ plays a significant role in modifying the energy and momentum profiles of particle orbits.

Constraints on the Lorentz symmetry breaking parameter $l$ have been derived from various astrophysical observations. From the analysis of the periastron precession of the star S2 orbiting around Sgr A*, the bound $-0.185022 \leq l \leq 0.0609$ was established in \cite{nosso}. This contrasts with the much tighter constraint $-6.1\times 10^{-13} \leq l \leq 2.8\times 10^{-14}$ obtained in \cite{KR} using measurements of the Shapiro time-delay effect. Additionally, based on limits imposed by the Event Horizon Telescope (EHT) on the shadow radius of Sgr A*, a broader range of admissible values for the spontaneous Lorentz symmetry breaking parameter was found: $-4.59\times 10^{-3} \leq l \leq 1.24\times 10^{-1}$ \cite{nosso}. In this work, we adopt representative non-zero values of the LV parameter, specifically $l = -0.1$ and $l = 0.1$. These values enable a meaningful exploration of the influence of Lorentz symmetry breaking on black hole accretion dynamics \footnote{ $l=-0.1$ is not within the numerical range obtained in the others in \cite{nosso} .  We adopt this value only to identify any modification caused by the $l$ parameter.}.

In the absence of a cosmological constant, the solution exhibits notable physical features that can be directly compared to the standard Schwarzschild solution.  The line element is given by:
\begin{equation}
    ds^2 = -A(r) dt^2 +A(r)^{-1}dr^2 + r^2\left(d\theta^2 + \sin^2 \theta d\phi^2\right)\,.
    \label{eq:dsgeral}
\end{equation}
with
\begin{equation}
A(r)=\left(\frac{1}{1 - l} - \frac{2M}{r}\right) \label{KR0}
\end{equation}
where $M$ is the mass parameter and $l$ is the LV parameter in KR gravity. Its presence modifies the geometry without altering its essential black hole character. Indeed, the (single) event horizon for this solution is located at $r_h = 2(1 - l)M$, thus corresponding to a black hole for every value of $l<1$.

The main aim of this article is to investigate the phenomenon of fluid accretion onto a KR black hole described by the line element (\ref{eq:dsgeral}) and (\ref{KR0}), with the fluid dynamics governed (mainly) by its equation of state. As a first step, in Sec.\,\ref{sec1} we consider Bondi-like accretion, generalized to the relativistic regime by Michel. The process of accretion of matter in KR gravity black holes is discussed in Sec. \ref{sec3}. In Sec.\,\ref{Sec3}  we describe such a process using a Hamiltonian dynamical system framework. The analysis of critical points for different types of isothermal fluids is presented in Sec.\,\ref{seciso}, where we explore the system’s evolution and identify sonic transitions. In all these investigations, the influence of the LV parameter is systematically analyzed.
Furthermore, in Sec.\,\ref{sec6}, we study the behavior of polytropic fluids within the Hamiltonian formalism, focusing on the structure and location of critical points. Finally, Sec.\,\ref{Sec:Conclusion} provides a summary of our results and concluding remarks.
Throughout this work, we adopt geometrized units ($G = 1$ and $c = 1$) and use the metric signature $(-,+,+,+)$.

%%%%%%%%%%%%%%%%%%%%%%%%%%%%%%%%%%%%%%%%%%%%%%%%%%%%%%%%%%%%%%%
\section{Dynamical equations of mass accretion} \label{sec1}
%%%%%%%%%%%%%%%%%%%%%%%%%%%%%%%%%%%%%%%%%%%%%%%%%%%%%%%%%%%%%%%

%The lagrangean that describes a test particle\label{•} in GR is  $\mathcal{L}=\frac{1}{2}g_{\mu\nu}\dot{x}^{\mu}\dot{x}^{\nu}$, where the overdot denotes differentiation with respect to an affine parameter. When a particle moves along timelike geodesics, it must satisfy the condition $\mathcal{L}=-1$, and therefore we have \cite{Inverno}
%\begin{equation}
%g_{\mu\nu}\dot{x}^{\mu}\dot{x}^{\nu}=-1.
%\label{gmn}
%\end{equation}

In this section we shall establish various properties of fluid accretion, including the behavior of the energy density, the radial velocity of the flow, the speed of sound, and the rate of matter accretion onto the black hole, whose general spacetime structure is described by \eqref{eq:dsgeral}. For this purpose, the accreting matter is modelled as a perfect, isotropic, and non-homogeneous fluid, characterized by
\begin{eqnarray}
T_{\mu\nu}=(\rho+p)u_\mu u_\nu +p g_{\mu\nu} 
\end{eqnarray}
where $\rho$ and $p$ denote the energy density and pressure of the fluid, respectively. The four-velocity is defined as $u^\mu = dx^\mu / d\tau = (u^t, u^r, 0, 0)$, where $\tau$ represents the proper time of the particles in the fluid. This four-velocity satisfies the normalization condition $u_\mu u^\mu = -1$, from which we obtain the relation
\begin{eqnarray}
u^t=\frac{\sqrt{(u^r)^2+A(r)}}{A(r)}\,.\label{ut}
\end{eqnarray}
If $u^r > 0$, the fluid is being ejected from the black hole, whereas $u^r < 0$ indicates fluid accretion. On the other hand, the sign of $u^t$ determines the temporal orientation of the flow: $u^t > 0$ corresponds to forward evolution in time, while $u^t < 0$ implies a backward progression. It is important to emphasize that the condition $u^t > 0$ ensures the causality of the process, meaning that the fluid can only fall into the black hole, as expected in a physically consistent scenario.

From the conservation law of the energy-momentum tensor, $\nabla_\mu T^{\mu\nu}=0$, we get
\begin{eqnarray}
(\rho+p)u^rr^2\sqrt{(u^r)^2+A(r)}=\mathcal{A}_0\,,\label{A1}
\end{eqnarray}
where $\mathcal{A}_0$ is an integration constant, and the fluid is assumed to obey an equation of state of the form $p = p(\rho)$. The relationship between the energy-momentum conservation law and the four-velocity is established through the projection $u_\mu \nabla_\nu T^{\mu\nu} = 0$, which yields the continuity equation. This equation can be solved by employing the normalization condition $u_\mu u^\mu = -1$ and the identity $\nabla_\mu g^{\mu\nu} = 0$, leading to
\begin{eqnarray}
r^2 u^r \exp\left[\int \frac{d\rho}{\rho+p}\right]=-\mathcal{A}_1\,.\label{A0}
\end{eqnarray}
Here, $\mathcal{A}_1$ denotes a constant of integration. Since the radial velocity satisfies $u^r < 0$ during accretion, the right-hand side of Eq. \eqref{A0} carries a negative sign. Dividing Eq. \eqref{A1} by \eqref{A0} yields
\begin{eqnarray}
(\rho+p)\sqrt{(u^r)^2+A(r)}\exp\left[-\int \frac{d\rho}{\rho+p}\right]=\mathcal{A}_2\,,\label{A3}
\end{eqnarray}
where $\mathcal{A}_2$ is a constant of integration. 

On the other hand, the mass flow equation is given by
\begin{eqnarray}
\nabla_\mu J^\nu\equiv 0\,,\label{DivJ}
\end{eqnarray}
where $J^\mu=nu^\mu$, with $n$ being the baryonic density number. Considering the equatorial plane and taking $n=\rho$, we obtain from Eq. \eqref{DivJ} that
\begin{eqnarray}
\rho u^r r^2= \mathcal{A}_3 \,, \label{A2}
\end{eqnarray}
and $\mathcal{A}_3$ is a constant of integration. Dividing Eq. \eqref{A1} by \eqref{A2} we get 
\begin{eqnarray}
\frac{(\rho+p)}{\rho}\sqrt{(u^r)^2+A(r)}=\mathcal{A}_4 \label{A4}
\end{eqnarray}
with $\mathcal{A}_4$ a constant.

%%%%%%%%%%%%%%%%%%%%%%%%%%%%%%%%%%%%%%%%%%%%%%%%%%%%%%%%%%%%%%%
\subsection{Critical parameters}
%%%%%%%%%%%%%%%%%%%%%%%%%%%%%%%%%%%%%%%%%%%%%%%%%%%%%%%%%%%%%%%

As the fluid accelerates toward the black hole under the influence of its strong gravitational field, it eventually reaches a location where its velocity matches the local speed of sound. This location is known as the sonic point (critical point), and depending on the solution, multiple sonic points may exist. At the sonic point, the accretion flow attains its maximum rate. Following the procedure outlined in \cite{Ujal, Jamil3}, and starting from Eqs. \eqref{A2} and \eqref{A4}, we derive the following differential equation:
\begin{eqnarray}
&&\left(v^2-\frac{(u^r)^2}{(u^r)^2+A(r)}\right)\frac{du}{u}\nonumber\\
&&+\left(\frac{2}{r}v^2-\frac{A'(r)}{2\left((u^r)^2+A(r)\right)}\right)dr=0\,. \label{eqCP}
\end{eqnarray}
where the prime denotes a differentiation with respect to $r$ and
\begin{eqnarray}
v^2=\frac{d\,{\rm ln}(\rho+p)}{d\,{\rm ln\rho}}-1\,,\
\end{eqnarray}
is interpreted as the speed of sound in the medium, i.e. $v^2=v^2_s=\frac{dp}{d\rho}$ \cite{Babichev2}. 
The critical points are determined by simultaneously setting the expressions within the parentheses in Eq. \eqref{eqCP} to zero, thereby decoupling the system and yielding
\begin{eqnarray}
u^2_c=\frac{1}{4}r_c A'(r_c)\,,\,\,\,\,\,\, v^2_c=\frac{r_c A'(r_c)}{4A(r_c)+r_cA'(r_c)} \label{uVc}
\end{eqnarray}
where $u_c = u^r_c$ denotes the critical fluid velocity evaluated at the critical radius $r = r_c$. Physically acceptable solutions require that both $u_c^2 > 0$ and $v_c^2 > 0$ hold true, and the set of sonic points obtained by solving Eq. \eqref{uVc} is given by $(r_c, \pm u^r_c)$ \cite{Ujal, Jamil3}. Consequently, the fluid velocity must increase monotonically and pass smoothly through the critical point during the accretion process.

%%%%%%%%%%%%%%%%%%%%%%%%%%%%%%%%%%%%%%%%%%%%%%%%%%%%%%%%%%%%%%%
\subsection{Mass accretion rate} \label{Sec2}
%%%%%%%%%%%%%%%%%%%%%%%%%%%%%%%%%%%%%%%%%%%%%%%%%%%%%%%%%%%%%%%

In realistic astrophysical systems, the mass of the black hole evolves over time. While accretion leads to an increase in mass, Hawking radiation acts to decrease it. The rate of change of the black hole mass can be determined by integrating the flux of the fluid across the two-dimensional event horizon surface, yielding
\begin{eqnarray}
\dot{M}&=&-\int \sqrt{-g} (p+\rho)u_tu^r d\theta d\phi \nonumber\\
&=&4\pi \mathcal{A}_2(p_\infty+\rho_\infty)\sqrt{A(r_\infty)}M^2
\end{eqnarray}
where we have used Eqs. \eqref{A0} and \eqref{A3}. It is important to exercise caution with the term $\sqrt{A(r_\infty)}$ since, in the scenario under consideration, the metric does not approach the Minkowski form asymptotically. Therefore, the accretion rate is given by
\begin{eqnarray}
\dot{M}_{ac}=4\pi \mathcal{A}_2  (p+\rho)\sqrt{A(r_\infty)} M^2\,,\label{Mac}
\end{eqnarray}
where the dot denotes the derivative with respect to time. As $\dot{M}$ is time-dependent, the increase $(\rho + p > 0)$ or decrease $(\rho + p < 0)$ of the black hole mass is determined by the nature of the fluid accreting onto it.

In the presence of evaporation due to Hawking radiation, the corresponding mass loss rate is given by \cite{Cline, Manuela} 
\begin{eqnarray}
\dot{M}_{ev}=-\frac{\alpha}{M^2}\,,\label{mev}
\end{eqnarray}
where $\alpha$ is a characteristic constant associated with Hawking radiation. It is important to note that the evaporation process is independent of the fluid properties. Consequently, the total rate of change of the black hole mass is given by $\dot{M} = \dot{M}_{ac} + \dot{M}_{ev}$ \cite{Ujal}. In this expression, $\dot{M} < 0$ holds when 
\[
M^4 < \frac{\alpha}{4\pi \mathcal{A}_2 (p + \rho) \sqrt{A(r_\infty)}},
\]
and $\dot{M} > 0$ when the inequality is reversed. These conditions apply for both ordinary fluids and quintessence, whereas for phantom energy the black hole mass always decreases \cite{Babichev, Babichev2}. A similar analysis was previously conducted in \cite{Ujal}. Since our focus is solely on black hole mass accretion under LV, in what follows we set $\alpha = 0$ in Eq.  \eqref{mev}.

%%%%%%%%%%%%%%%%%%%%%%%%%%%%%%%%%%%%%%%%%%%%%%%%%%%%%%%%%%%%%%%
\section{Accretion of matter by black hole with LV}\label{sec3}
%%%%%%%%%%%%%%%%%%%%%%%%%%%%%%%%%%%%%%%%%%%%%%%%%%%%%%%%%%%%%%%

We consider an isothermal fluid model governed by the equation of state $p = \omega \rho(r)$, where the parameter $\omega$ characterizes the type of matter. This model is physically well motivated, as it assumes that the fluid remains at constant temperature during thermodynamic processes. Consequently, the speed of sound remains constant and is given by $v_s^2 = \omega = \mathrm{d}p / \mathrm{d}\rho$.

We can then solve the system of equations formed by Eqs. \eqref{A1} and \eqref{A4} to obtain
\begin{eqnarray}
u(r)=\pm \frac{\sqrt{\mathcal{A}^2_4-A(r)-2\omega A(r)-\omega^2 A(r)}}{(1+\omega)} \label{u}
\end{eqnarray}
and 
\begin{eqnarray}
\rho (r)=\pm \frac{\mathcal{A}_3 (\omega +1)}{r^2 \sqrt{\mathcal{A}^2_4-A(r)-2\omega A(r)-\omega^2 A(r)}}\,, \label{rho}
\end{eqnarray}
where we define the constant $\mathcal{A}_4 \equiv \mathcal{A}_0 / \mathcal{A}_3$. For the black hole with LV described by Eq. \eqref{KR0}, the radial velocity $u^r = u(r)$ and the fluid density $\rho = \rho(r)$ are obtained from Eqs. \eqref{u} and \eqref{rho}, and are given by
\begin{eqnarray}
u(r)=\pm\frac{1}{1+\omega}\sqrt{\mathcal{A}^2_4+\left(1+\omega\right)^2\left(\frac{1}{1-l}-\frac{2M}{r}\right)}\,,\label{uLV}
\end{eqnarray}
and 
\begin{eqnarray}
\rho(r)=\pm\frac{\mathcal{A}_3(1+\omega)}{r^2}\left[\mathcal{A}^2_4+\left(1+\omega\right)^2\left(\frac{1}{1-l}-\frac{2M}{r}\right)\right]^{-1/2}\,, \label{rhoLV}
\end{eqnarray}
respectively. Collecting all this information, the accretion rate of matter \eqref{Mac} takes the following form:
\begin{eqnarray}
\dot{M}_{ac}=\frac{4\pi (1+\omega) \mathcal{A}^2_3\mathcal{A}_4}{(1-l)r^2\sqrt{\mathcal{A}^2_4+\left(1+\omega\right)^2\left(\frac{1}{1-l}-\frac{2M}{r}\right)}}\,. \label{MacLV}
\end{eqnarray}

We now proceed to analyze the monotonic behavior of the fluid density and radial velocity, as well as the rate of change of the black hole mass, with respect to the LV parameter $l$. Since the equation of state is given by $p = \omega \rho(r)$, we can set $\mathcal{A}_3 = 1$ in all cases without loss of generality \cite{Bahamond}. Due to the symmetry in the expressions for $u(r)$ and $\rho(r)$, we adopt the convention $u(r) = |u(r)|$ and $\rho(r) = |\rho(r)|$ in the plots that follow.
Our analysis will focus on the following types of fluids: ultra-stiff fluid ($\omega = 1$), ultra-relativistic fluid ($\omega = 1/2$), and radiation ($\omega = 1/3$).

\begin{figure*}[ht!]
\centering
\subfigure[] 
{\label{roa}\includegraphics[width=8.5cm]{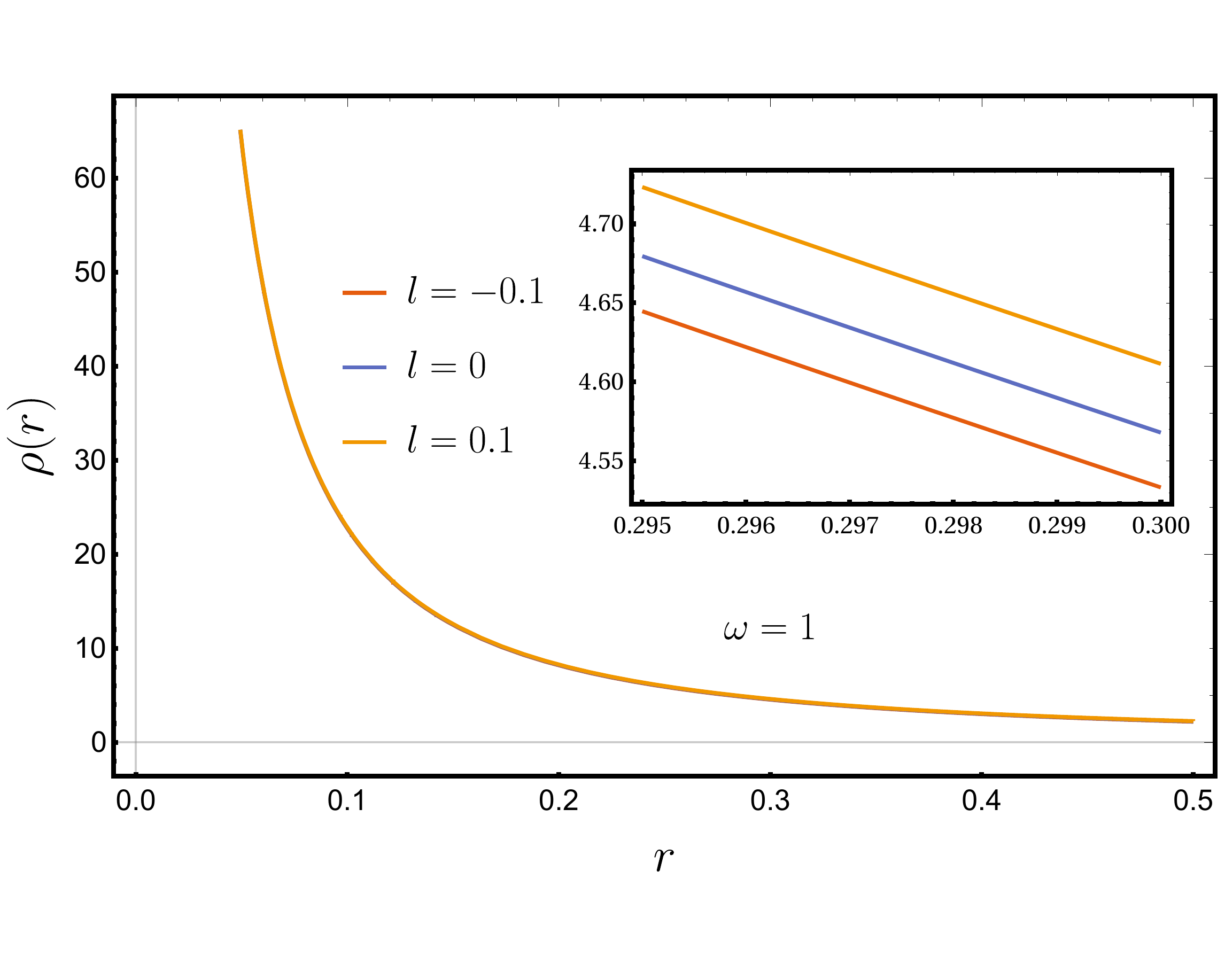} }
\subfigure[] 
{\label{rob}\includegraphics[width=8.5cm]{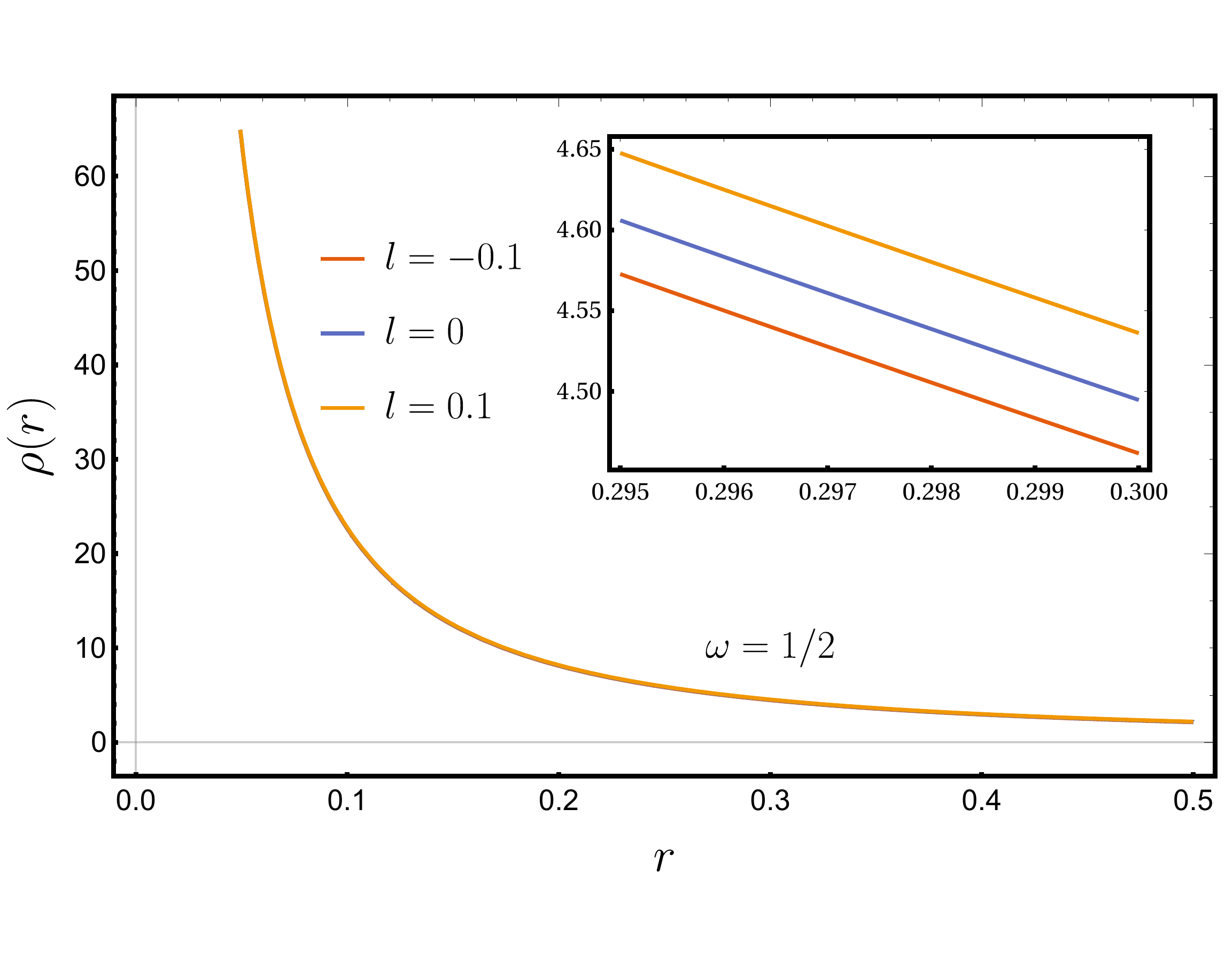} }
\subfigure[]
{\label{roc}\includegraphics[width=8.5cm]{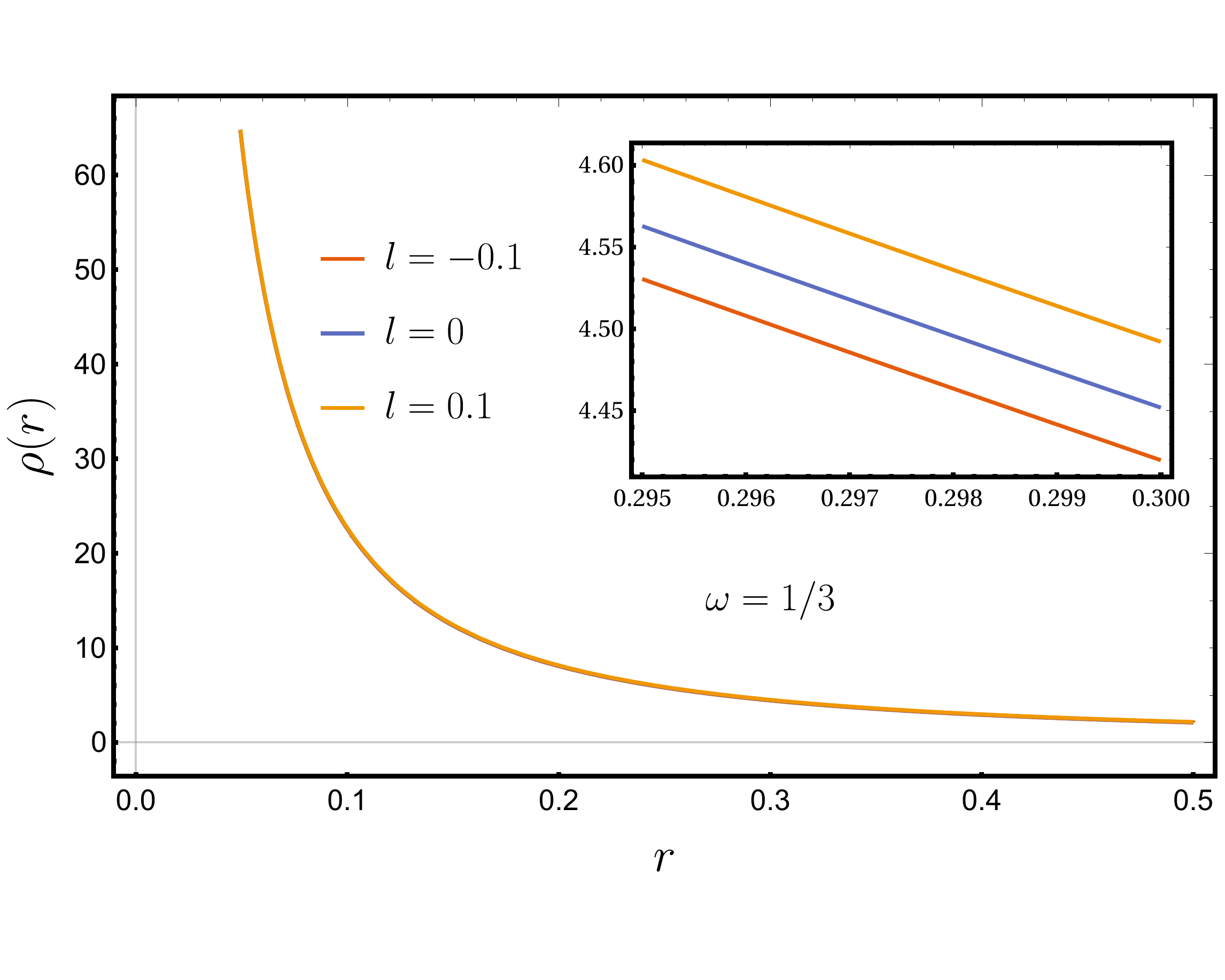} }
%\subfigure[]
%{\label{rod}\includegraphics[width=8.5cm]{rho14.pdf} }
%\hspace{0.1cm}
\caption{The energy density of Eq. \eqref{rhoLV} for ultra-stiff ($\omega=1$, top left), ultra-relativist ($\omega=1/2$, top right) and radiation ($\omega=1/3)$, bottom) with $l>0$ and $l<0$, as compared to Schwarzschild solution. Here we made $\mathcal{A}_3=\mathcal{A}_4=1$.}\label{Fig1}
\end{figure*}

The influence of the LV parameter $l$ on the behavior of the fluid energy density, as given by Eq. \eqref{rhoLV}, is illustrated in Fig.\,\ref{Fig1} for each specific type of matter. In all cases considered, corresponding to different values of the equation of state parameter $\omega$, the energy density increases as the fluid approaches the black hole. We observe that for $l > 0$ ($l<0$), the energy density is enhanced (suppressed) as compared to the Schwarzschild case. Additionally, it is observed that the fluid energy density asymptotically vanishes at spatial infinity for all values of $\omega$.
  
\begin{figure*}[ht!]
\centering
\subfigure[$\omega=1$] 
{\label{ua}\includegraphics[width=8.5cm]{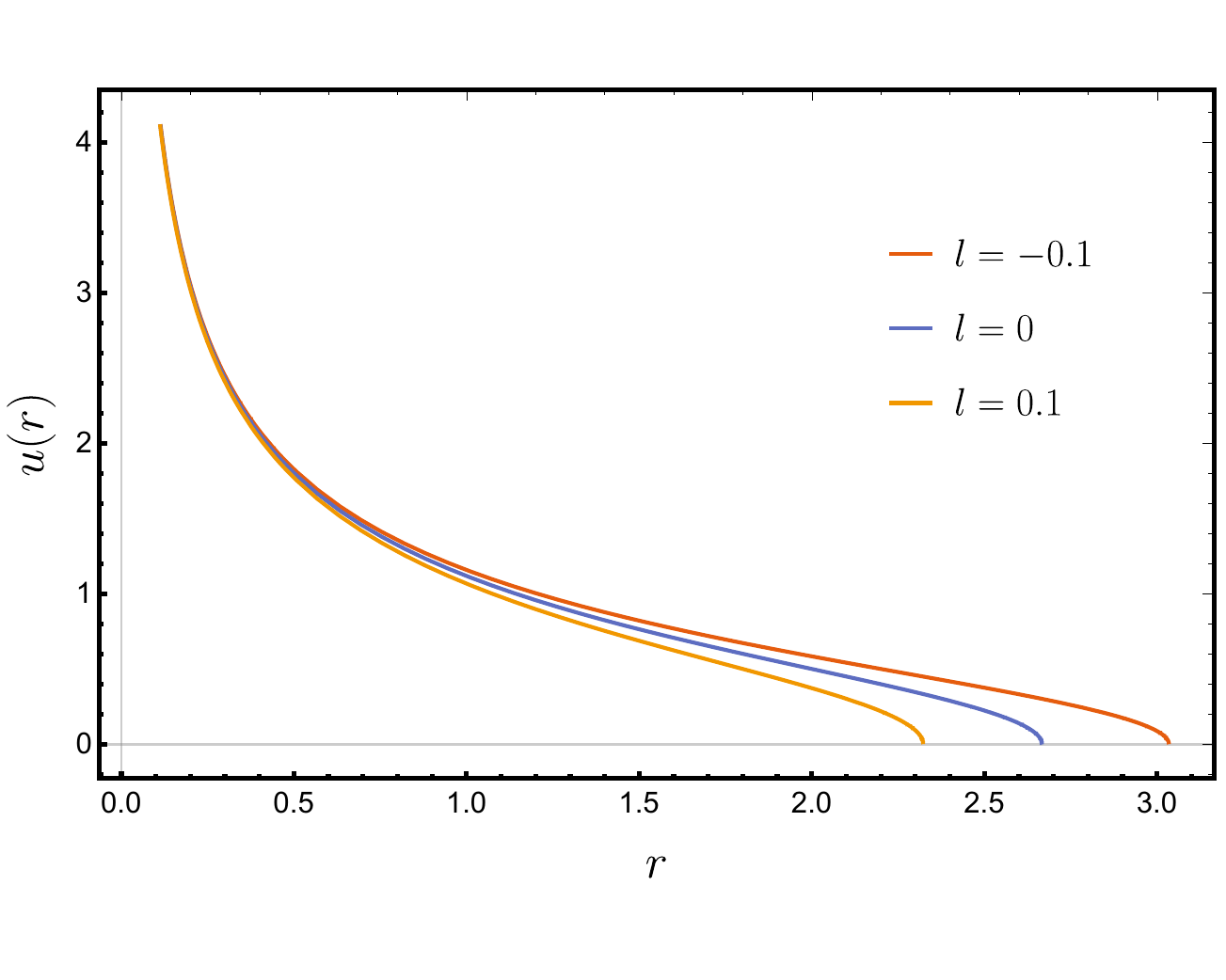} }
\subfigure[$\omega=1/2$] 
{\label{ub}\includegraphics[width=8.5cm]{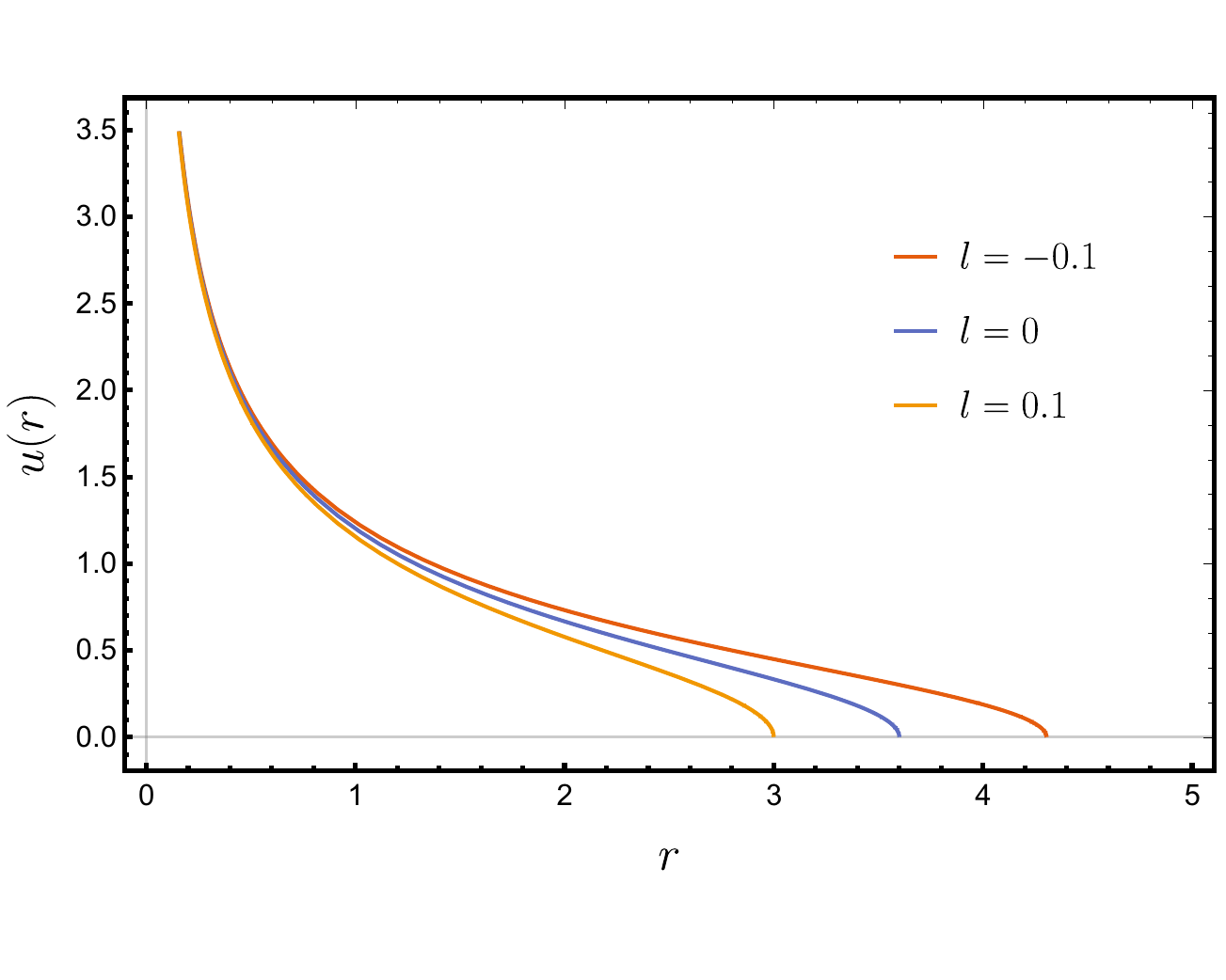} }
\subfigure[$\omega=1/3$]
{\label{uc}\includegraphics[width=8.5cm]{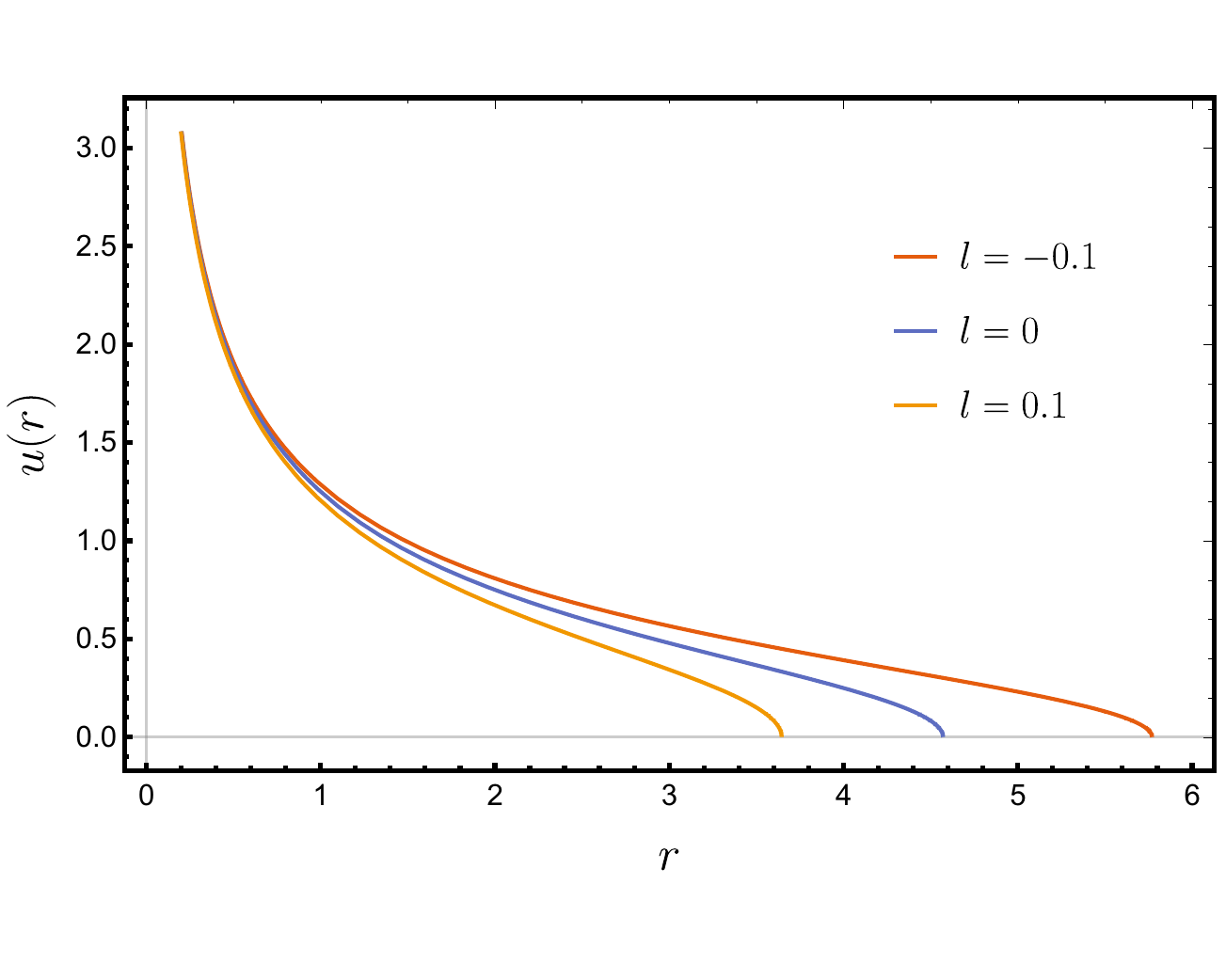} }
\caption{The radial velocity of Eq. \eqref{uLV} for ultra-stiff ($\omega=1$, top left), ultra-relativist ($\omega=1/2$, top right) and radiation ($\omega=1/3)$, bottom) with $l>0$ and $l<0$, as compared to Schwarzschild solution. Here we made $\mathcal{A}_4=1$. }\label{Fig2}
\end{figure*}

The radial velocity, given by Eq. \eqref{uLV}, is plotted as a function of the radial coordinate $r$ in Fig.\,\ref{Fig2}, illustrating the influence of the LV parameter $l$ for various values of the equation of state parameter $\omega$. As shown in this figure, the radial velocity of the fluid increases as it approaches the black hole, eventually reaching the speed of sound and subsequently attaining supersonic values due to the intense gravitational field. For the isothermal fluids considered in this analysis, the LV parameter $l$ modifies the behavior of the radial velocity such that, for $l < 0$, the fluid crosses the sonic point farther from the black hole, whereas for $l > 0$, the transition to supersonic flow occurs closer to the event horizon.

\begin{figure*}[ht!]
\centering
\subfigure[] 
{\label{Ma}\includegraphics[width=8.5cm]{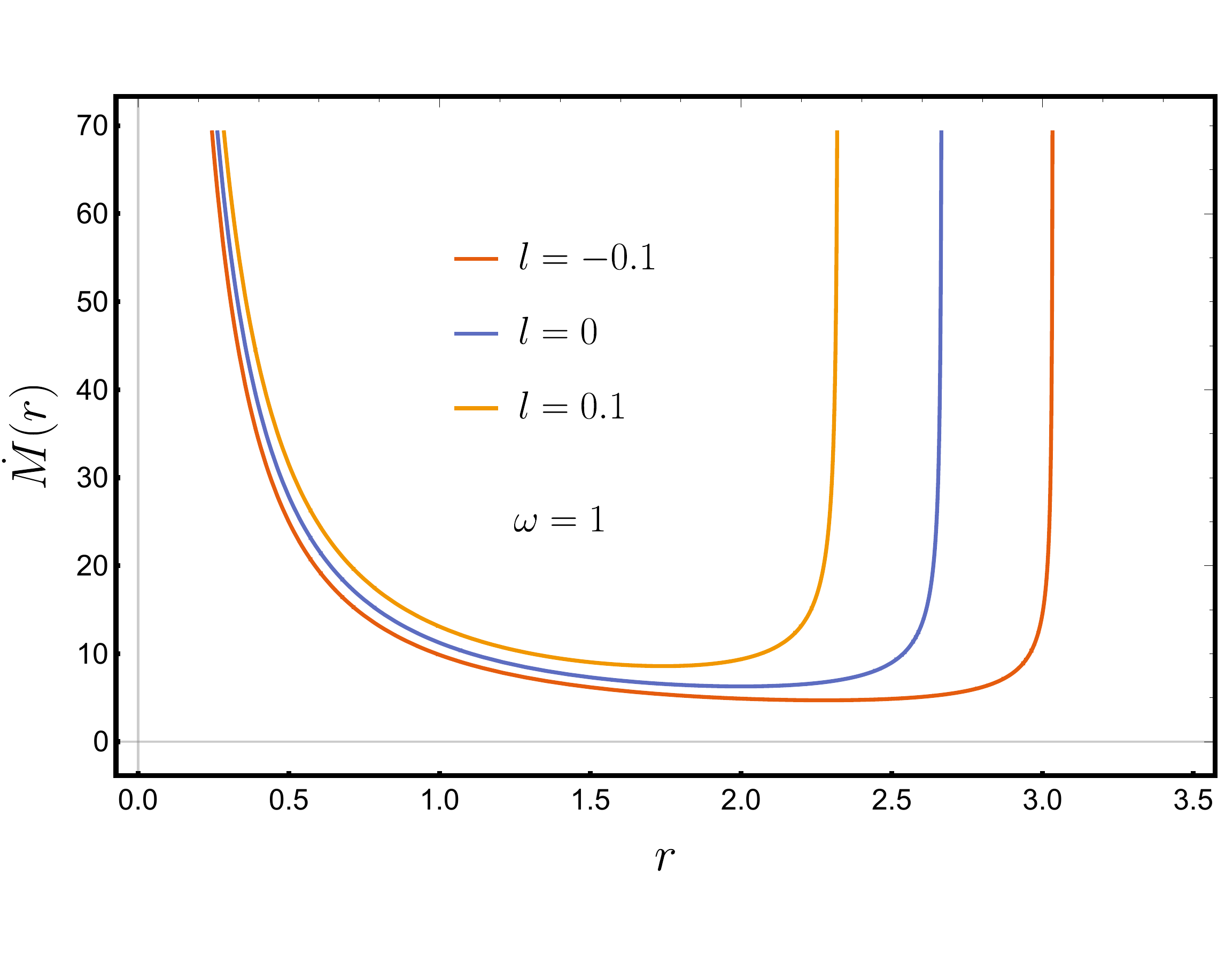} }
\subfigure[] 
{\label{Mb}\includegraphics[width=8.5cm]{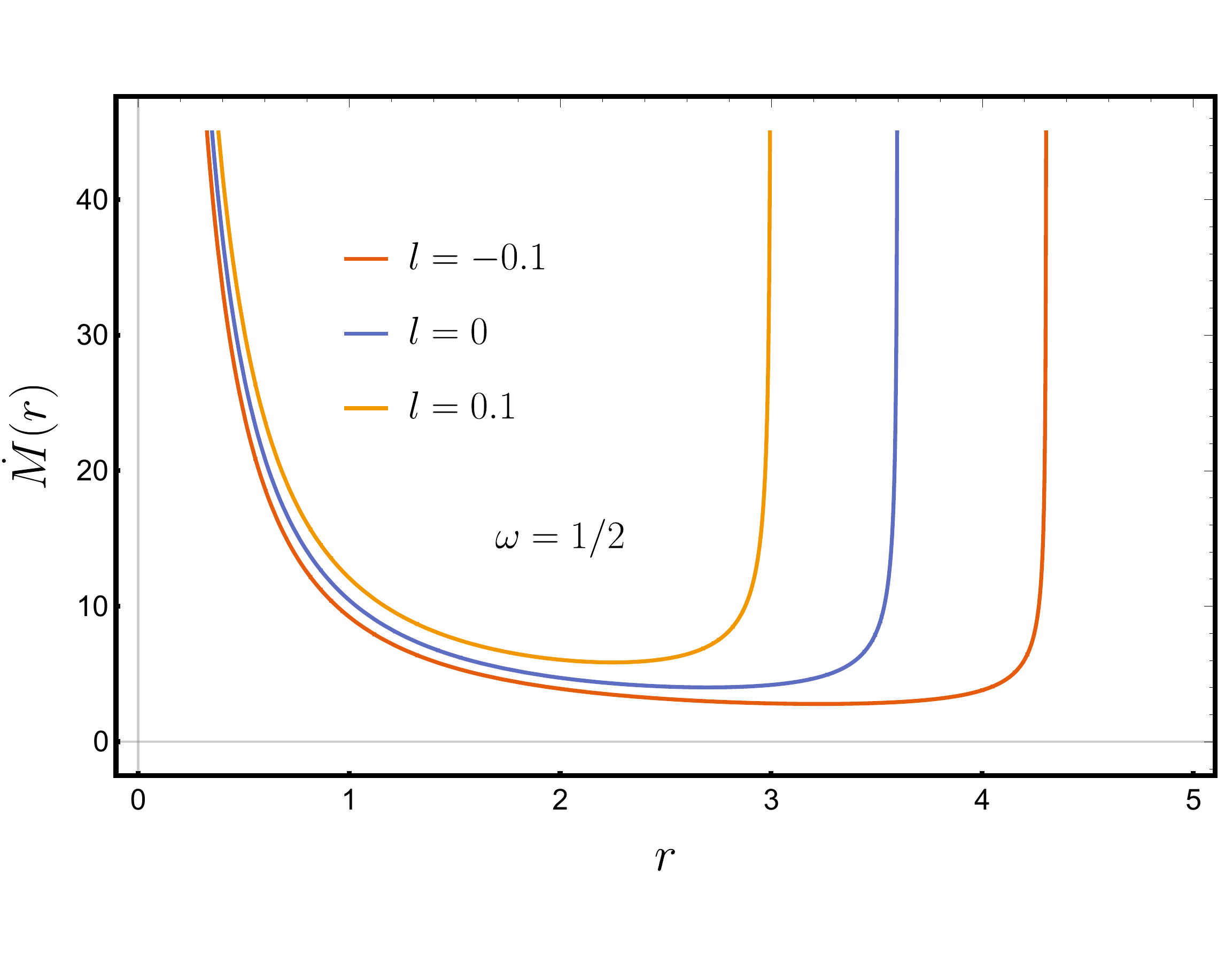} }
\subfigure[]
{\label{Mc}\includegraphics[width=8.5cm]{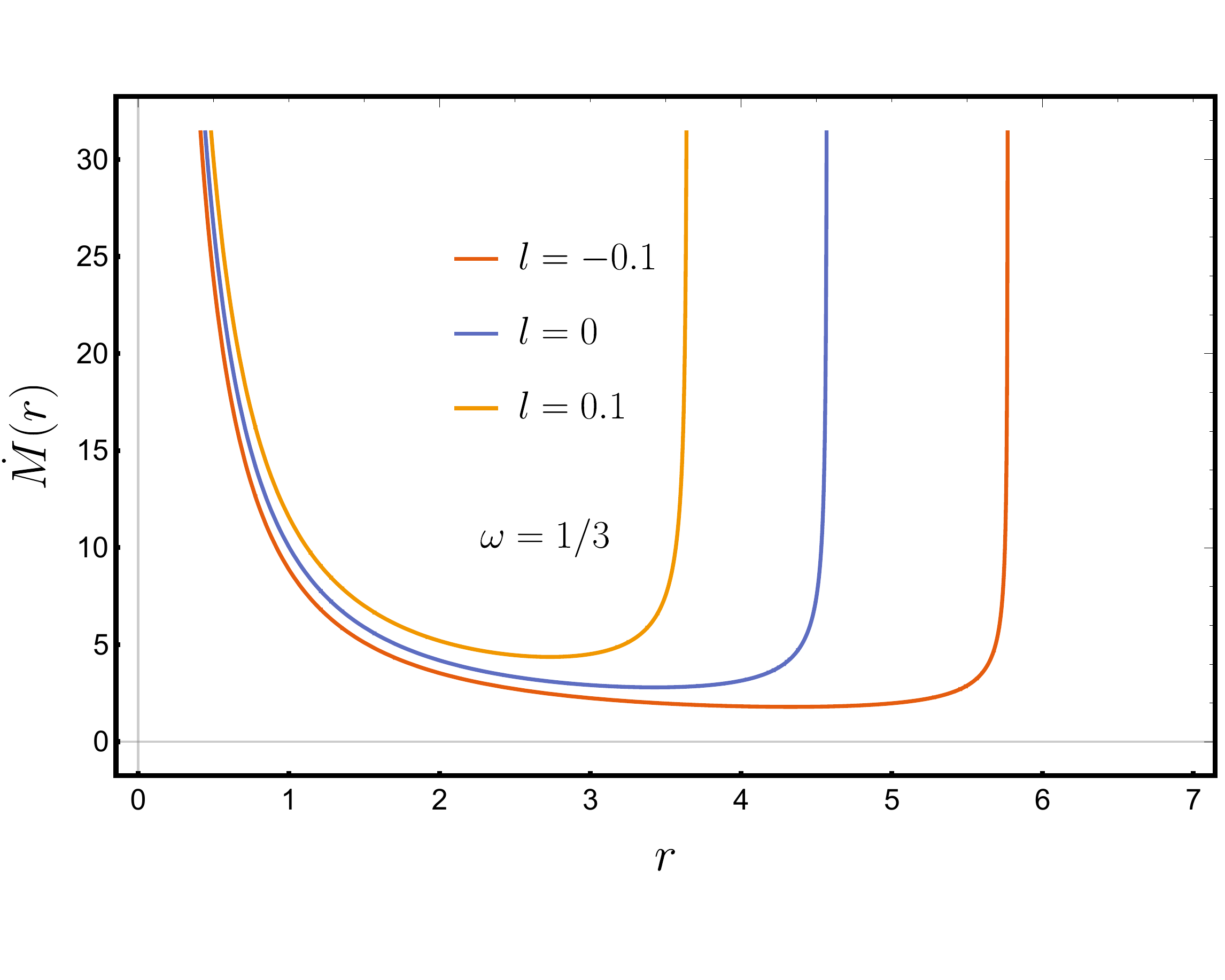} }
%\subfigure[]
%{\label{ud}\includegraphics[width=8.5cm]{M14.pdf} }
%\hspace{0.1cm}
\caption{The mass accretion rate of Eq. \eqref{MacLV} for ultra-stiff ($\omega=1$, top left), ultra-relativist ($\omega=1/2$, top right) and radiation ($\omega=1/3)$, bottom) with $l>0$ and $l<0$, as compared to Schwarzschild solution. Here we made $\mathcal{A}_3=\mathcal{A}_4=1$.
}\label{Fig3}
\end{figure*}

The accretion rate, given by Eq. \eqref{MacLV}, is depicted in Fig.\,\ref{Fig3} for various isothermal fluids under the influence of the LV parameter. It is evident that the accretion rate increases in the vicinity of the black hole for all values of the equation of state parameter $\omega$ considered here. Moreover, the accretion rate is enhanced (suppressed) for positive (negative) values of $l$. In all cases, however, the black hole experiences a net increase in mass due to the continuous accretion of matter characterized by $\omega$. Similar qualitative behavior has been reported in the context of black holes coupled to nonlinear electrodynamics \cite{Mustafa} and Horndeski/Galileon black holes \cite{Nozari}.

The analysis conducted thus far provides insights into how the fluid velocity increases as it is accelerated by the gravitational field of the black hole. However, to perform a more precise investigation of the critical points, we will construct a dynamical system that enables the study of the fluid's adiabatic behavior.

%%%%%%%%%%%%%%%%%%%%%%%%%%%%%%%%%%%%%%%%%%%%%%%%%%%%%%%%%%%%%%%
\section{Michel-type accretion via Hamiltonian systems}\label{Sec3}
%%%%%%%%%%%%%%%%%%%%%%%%%%%%%%%%%%%%%%%%%%%%%%%%%%%%%%%%%%%%%%%

In this section, we analyze the isothermal accretion of matter by employing the Hamiltonian dynamical system approach developed in \cite{Amed2, ednaldo1}. We begin with two fundamental equations that govern the process of accretion (or ejection) of matter by black holes. The first is the conservation equation for the matter flux, given by Eq. \eqref{DivJ}, which, when restricted to the equatorial plane and integrated, yields
\begin{eqnarray}
	r^2 n u = c_1\,,\label{c1}
\end{eqnarray}
where $c_1$ is a constant of integration.

According to Bernoulli's theorem in relativistic hydrodynamics, there exists a conserved quantity along the fluid flow in the co-moving frame. This is expressed through the second fundamental equation, namely the conservation of energy,
\begin{eqnarray}
	u^\nu\nabla_\nu\left[hu_\mu\zeta^\mu\right] \equiv 0\,,\label{Econserv}
\end{eqnarray}
where $h$ is the specific enthalpy and $\zeta^\mu$ denotes the timelike Killing vector associated with the static symmetry of the spacetime. Due to spherical symmetry, the Killing vector is given by $\zeta^\mu = [1,0,0,0]$, and using the relation $u_t = g_{tt} u^t$, from Eqs.\,\eqref{ut} and \eqref{Econserv}, we obtain
\begin{eqnarray}
	h\sqrt{A(r) + u^2} = c_2\,,\label{c2}
\end{eqnarray}
where $u = u^r$ and $c_2$ is a constant of integration.

The three-velocity of the fluid can be related to the temporal and radial components of the four-velocity via the line element restricted to the equatorial plane, that is
\begin{eqnarray}
	ds^2(\theta = \pi/2) = -\left( \sqrt{A(r)}\, dt \right)^2 + \left( \frac{dr}{\sqrt{A(r)}} \right)^2\,.
\end{eqnarray}
From this, the three-velocity of the fluid is defined by
\begin{eqnarray}
	 {\rm v}=\frac{dr}{A(r)dt}.
\end{eqnarray}
Using the definitions $u = dr/d\tau$, $u^t = dt/d\tau$, and $u_t = -A(r) u^t$, along with Eq.\,\eqref{ut}, we find
\begin{eqnarray}
	{\rm v}^2 = \left( \frac{u}{A(r) u^t} \right)^2 = \frac{u^2}{u_t^2} = \frac{u^2}{A(r) + u^2}\,,
\end{eqnarray}
which leads to the expressions
\begin{eqnarray}
	u^2 = \frac{A(r)\, {\rm v}^2}{1 - {\rm v}^2}\,, \quad \text{and} \quad u_t^2 = \frac{A(r)}{1 - {\rm v}^2}\,.\label{rmv}
\end{eqnarray}
Substituting these relations into Eq.\,\eqref{c1}, we obtain
\begin{eqnarray}
	\frac{r^4 n^2 {\rm v}^2 A(r)}{1 - {\rm v}^2} = c_1^2\,,\label{C12}
\end{eqnarray}
where we recall that $c_1$ is the integration constant from the mass flux conservation.

Finally, the conservation associated with the timelike Killing vector implies that the square of Eq.\,\eqref{c2} remains conserved, which allows us to define the Hamiltonian of the system as
\begin{eqnarray}
	\mathcal{H}(r, {\rm v}) = \frac{h(r, {\rm v})^2 A(r)}{1 - {\rm v}^2}\,.\label{Hlt}
\end{eqnarray}

Now, establishing the dynamical system from the Hamiltonian \eqref{Hlt}, we write
\begin{eqnarray}
	\frac{dr}{dt} = \frac{\partial \mathcal{H}}{\partial {\rm v}}\,, \quad \quad \frac{d{\rm v}}{dt} = -\frac{\partial \mathcal{H}}{\partial r}\,.
\end{eqnarray}
Following the procedure outlined in \cite{Amed2}, the dynamical system takes the form
\begin{eqnarray}
	\dot{r} &=& \frac{2 A(r) h^2}{{\rm v} (1 - {\rm v}^2)^2} \left( {\rm v}^2 - v_s^2 \right)\,, \label{dotr} \\
	\dot{{\rm v}} &=& -\frac{h^2}{r (1 - {\rm v}^2)} \left[ r \frac{d A(r)}{dr} (1 - v_s^2) - 4 A(r) v_s^2 \right]\,. \label{dotv}
\end{eqnarray}

We focus on the physically relevant critical point of the dynamical system, characterized by the conditions
\begin{eqnarray}
	{\rm v}_c = v_{s_c}\,, \quad \quad 4 A(r_c) v_{s_c}^2 = r_c \frac{d A(r_c)}{dr_c} \left( 1 - v_{s_c}^2 \right)\,.
\end{eqnarray}
From these, the speed of sound at the critical radius $r_c$ can be expressed as
\begin{eqnarray}
	v_{s_c}^2 = \frac{r_c A'(r_c)}{r_c A'(r_c) + 4 A(r_c)}\,, \label{vsomcrit}
\end{eqnarray}
which coincides with the second equation in Eq. \eqref{uVc}.

%%%%%%%%%%%%%%%%%%%%%%%%%%%%%%%%%%%%%%%%%%%%%%%%%%%%%%%%%%%%%%%
\section{Isotermal Fluid} \label{seciso}
%%%%%%%%%%%%%%%%%%%%%%%%%%%%%%%%%%%%%%%%%%%%%%%%%%%%%%%%%%%%%%%

Now, we consider an adiabatic perfect fluid, that is, a fluid which does not exchange heat with its surroundings. Additionally, we assume the fluid to be isentropic, meaning that its entropy remains constant. Consequently, the thermodynamic quantities are expressed as
\begin{eqnarray}
	dp = n \, dh \quad \text{and} \quad d\rho = h \, dn\,,
	\label{termo}
\end{eqnarray}
from which, dividing $dp$ by $d\rho$, the speed of sound $v_s^2$ is obtained as
\begin{eqnarray}
	v_s^2 = \frac{dp}{d\rho} = \frac{d \ln h}{d \ln n}\,.
	\label{som2}
\end{eqnarray}
Adopting the physical model where $v_s^2 = \omega$, it follows that $p = \omega \rho$. The enthalpy is defined as \cite{Amed2}
\begin{eqnarray}
	h = \frac{\rho + p}{n}.
\end{eqnarray}
Using \eqref{termo} and integrating, we find
\begin{eqnarray}
	\rho(n) = \frac{\rho_c}{n_c^{\omega+1}} \, n^{1+\omega},
\end{eqnarray}
and therefore, the enthalpy can be rewritten as
\begin{eqnarray}
	h = \frac{(1+\omega) \rho_c}{n_c^{\omega+1}} \, n^{\omega}.
	\label{hc}
\end{eqnarray}

To express the thermodynamic variables in terms of $r$ and ${\rm v}$, we write $n = n(r, {\rm v})$. Using Eq. \eqref{C12}, we have
\begin{eqnarray}
	n = \frac{c_1}{{\rm v} r^2} \sqrt{\frac{1 - {\rm v}^2}{A(r)}}\,.
	\label{n}
\end{eqnarray}
Substituting Eqs. \eqref{hc} and \eqref{n} into Eq. \eqref{Hlt} yields
\begin{eqnarray}
	\mathcal{H} = \mathcal{H}_0 \frac{1}{({\rm v} r^2)^{2\omega}} \left[ \frac{A(r)}{1 - {\rm v}^2} \right]^{1 - \omega}\,,
\end{eqnarray}
where
\begin{eqnarray}
	\mathcal{H}_0 = \left[ \frac{(1+\omega) \rho_c c_1^\omega}{n_c^{1+\omega}} \right]^2\,.
\end{eqnarray}
By applying a time reparametrization consistent with the Killing symmetry, $t \rightarrow \mathcal{H}_0 t$, the Hamiltonian rescales as $\mathcal{H} \rightarrow \mathcal{H} / \mathcal{H}_0$. Consequently, we redefine the Hamiltonian as
\begin{eqnarray}
	\mathcal{H}[r, {\rm v}] = \frac{1}{\left({\rm v} r^2\right)^{2\omega}} \left[ \frac{A(r)}{1 - {\rm v}^2} \right]^{1 - \omega}\,.
	\label{Hfinal}
\end{eqnarray}

We now apply the metric \eqref{KR0} to obtain the Hamiltonian \eqref{Hfinal} in the form
\begin{eqnarray}
	\mathcal{H}\left[r, l, {\rm v}\right] = \left(r^2 {\rm v}\right)^{-2\omega} \left[\frac{\frac{1}{1 - l} - \frac{2M}{r}}{1 - {\rm v}^2}\right]^{1 - \omega}\,.
	\label{Hfinal2}
\end{eqnarray}
The critical radius $r_c$ is found by setting \eqref{dotv} to zero using the metric \eqref{KR0}, resulting in
\begin{eqnarray}
	r_c = \frac{(1 - l) M (3\omega + 1)}{2 \omega}\,.
	\label{rc}
\end{eqnarray}
Note that $r_c$ depends explicitly on the LV parameter $l$ and the equation of state parameter $\omega$, implying that the critical point location varies with both fluid type and geometry. Substituting Eq. \eqref{rc} into Eq. \eqref{Hfinal2} and imposing ${\rm v}_c = v_s$ yields the critical Hamiltonian,
\begin{eqnarray}
	\mathcal{H}_c\left[l, \omega\right] &=& - \frac{\left[-48 l \omega - 16 l + 48 \omega + 16\right]^{\omega}}{(l - 1)(3\omega + 1)} \nonumber\\
	&& \times \left[\frac{(l - 1)^2 M^2 (3\omega + 1)^2}{\omega^{3/2}}\right]^{-2\omega}\,.
	\label{Hc}
\end{eqnarray}

In the following subsections, we analyze the dynamical system behavior for the different fluids introduced above.

%%%%%%%%%%%%%%%%%%%%%%%%%%%%%%%%%%%%%%%%%%%%%%%%%%%%%%%%%%%%%%%
\subsection{Ultra-stiff fluid}
%%%%%%%%%%%%%%%%%%%%%%%%%%%%%%%%%%%%%%%%%%%%%%%%%%%%%%%%%%%%%%%

The ultra-stiff fluid model, characterized by the equation of state parameter $\omega = 1$, represents the simplest case in our analysis. This scenario exhibits a universal behavior that is independent of the specific spherically symmetric metric considered. Such a result was previously obtained in \cite{Ahmed}, where the Hamiltonian reduces to
\begin{eqnarray}
	\mathcal{H} = \frac{1}{r^4 {\rm v}^2}
\end{eqnarray}
and the critical velocity is ${\rm v}_c = 1$. Notably, the dynamical system equations in this case do not admit any critical points.

\begin{figure*}[ht!]
\centering
\subfigure[$l>0$ with $r_h=1.8$] 
{\label{F4a}\includegraphics[width=8.5cm]{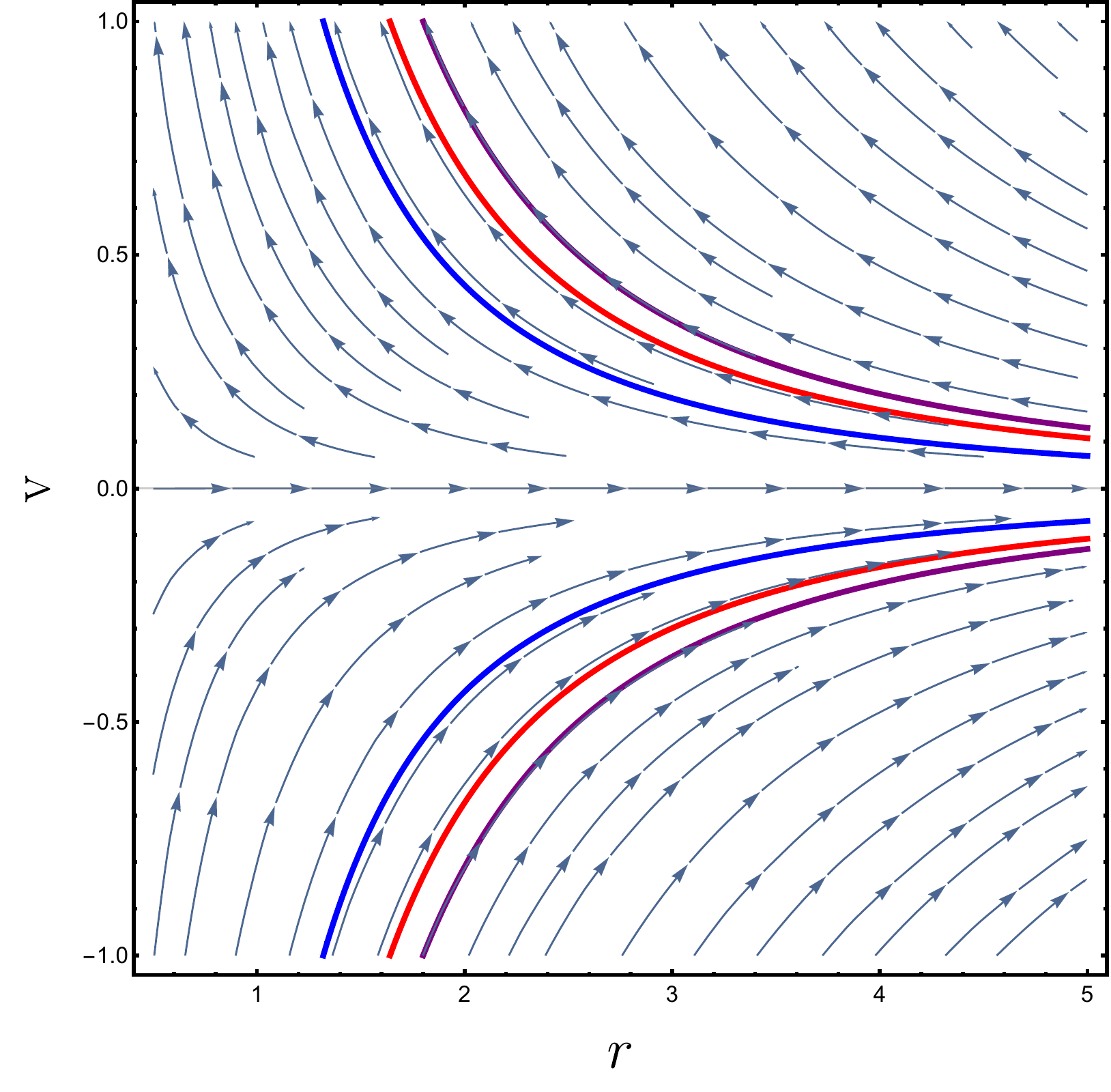} }
\hspace{0.1cm}
\subfigure[$l<0$ with $r_h=2.2$] 
{\label{F4b}\includegraphics[width=8.5cm]{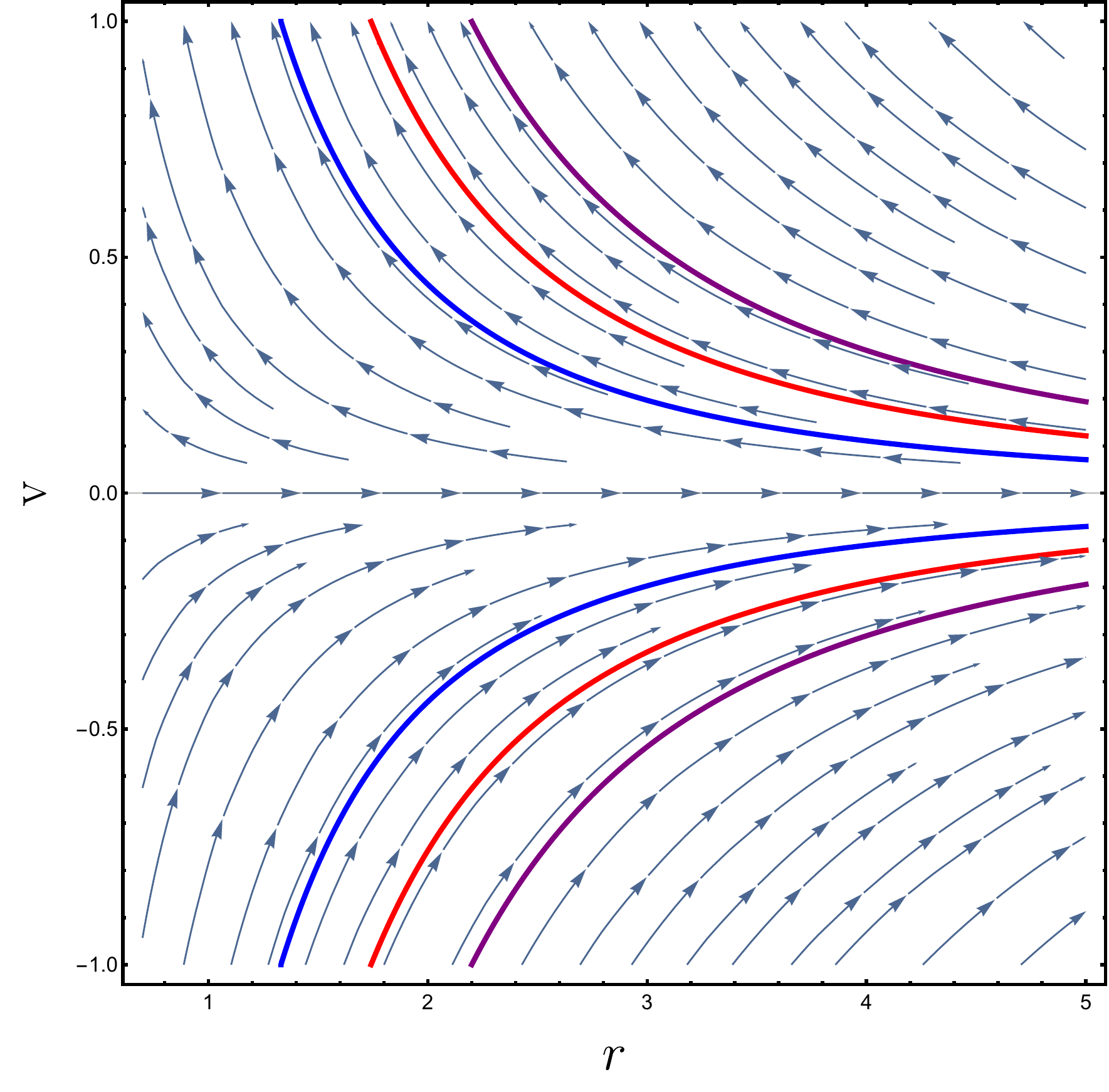} }
\hspace{0.1cm}
%\subfigure[]
%{\label{uc}\includegraphics[width=8.5cm]{M13.pdf} }
%\hspace{0.1cm}
%\subfigure[]
%{\label{ud}\includegraphics[width=8.5cm]{M14.pdf} }
%\hspace{0.1cm}
\caption{Representation of the phase space of the solution \eqref{KR0} for ultra-stiff fluid, $\omega=1$, and $l>0$ (left) and $l>0$ (right).}\label{Fig4}
\end{figure*}

The phase space diagram of the dynamical system for this fluid is depicted in Fig.\,\ref{Fig4}, illustrating two distinct cases: one with the LV parameter $l > 0$, where the horizon is located at $r_c = r_h = 1.8$, and another with $l < 0$, where the horizon is at $r_c = r_h = 2.2$. In both situations, the fluid flow displays a consistent pattern: the system's evolution either begins and ends with purely subsonic accretion (${\rm v} < 0$) or subsonic ejection (${\rm v} > 0$), represented by the red and blue curves, respectively.

Furthermore, the physically admissible flow occurs within the range $|{\rm v}| < 1$, confined between the lilac curves that correspond to the Hamiltonian's minimum values. While this qualitative behavior is global and independent of the value of $l$, it is important to note that the LV parameter does affect the location of the critical radius, which coincides with the event horizon, $r_c = r_h$.

%%%%%%%%%%%%%%%%%%%%%%%%%%%%%%%%%%%%%%%%%%%%%%%%%%%%%%%%%%%%%%%
\subsection{Ultra-relativistic fluid}
%%%%%%%%%%%%%%%%%%%%%%%%%%%%%%%%%%%%%%%%%%%%%%%%%%%%%%%%%%%%%%%

\begin{figure*}[ht!]
\centering
\subfigure[$l>0$ with $r_h=1.8$] 
{\label{F5a}\includegraphics[width=8.5cm]{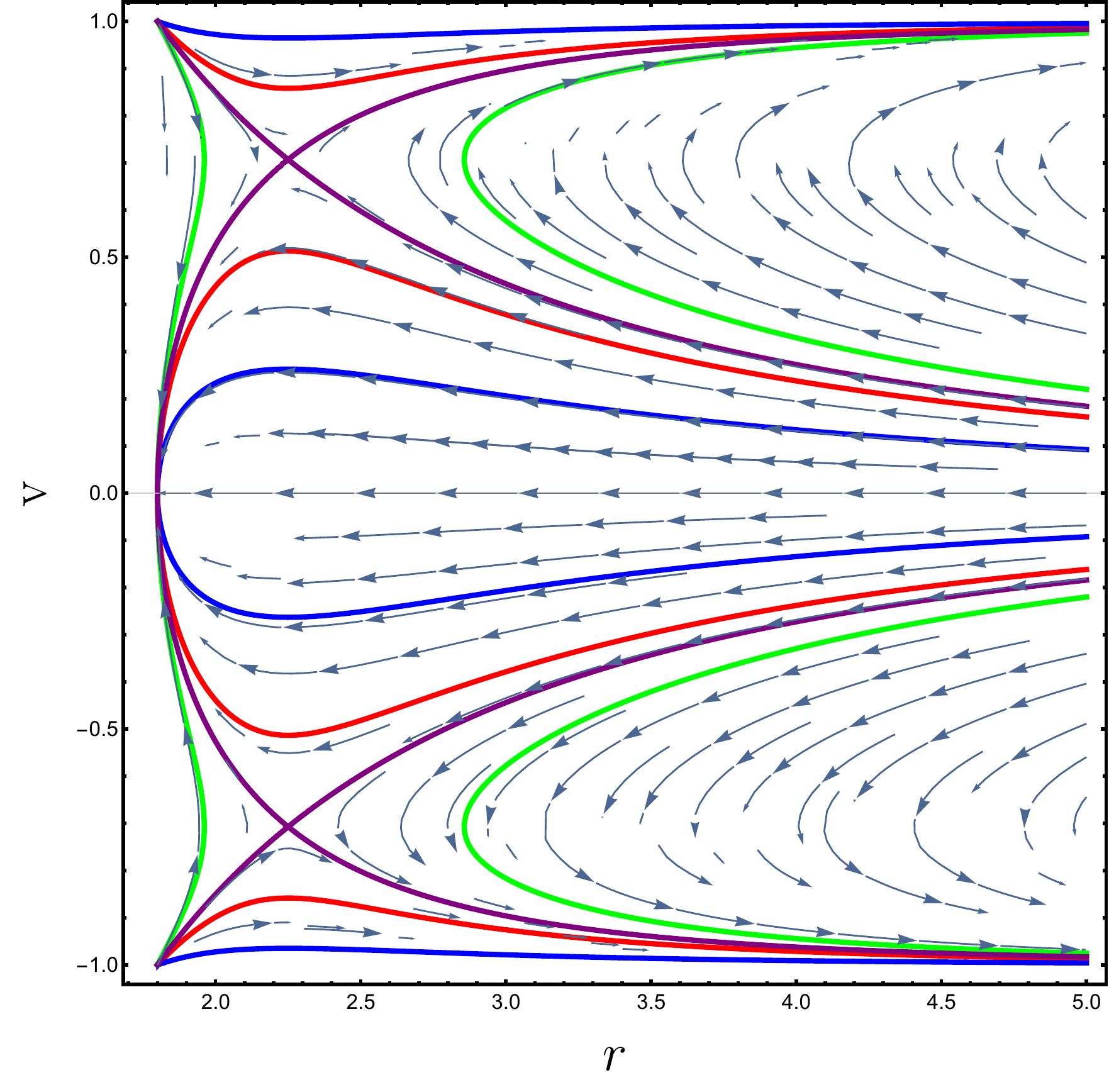} }
\hspace{0.1cm}
\subfigure[$l<0$ with $r_h=2.2$] 
{\label{F5b}\includegraphics[width=8.5cm]{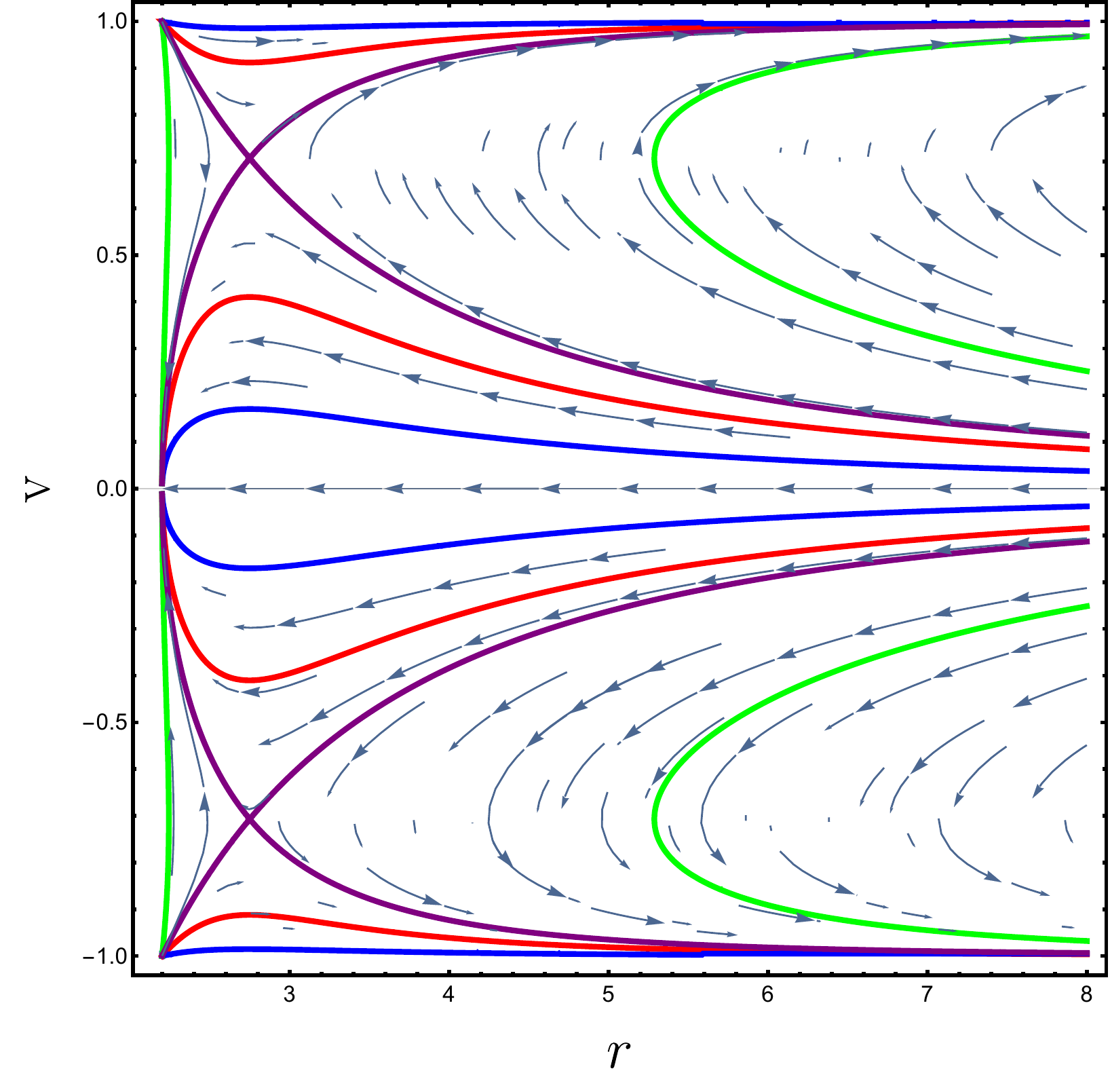} }
\hspace{0.1cm}
%\subfigure[]
%{\label{uc}\includegraphics[width=8.5cm]{M13.pdf} }
%\hspace{0.1cm}
%\subfigure[]
%{\label{ud}\includegraphics[width=8.5cm]{M14.pdf} }
%\hspace{0.1cm}
\caption{Representation of the phase space of the solution \eqref{KR0} for ultra-relativistic fluid, $\omega=1/2$, and $l>0$ (left) and $l>0$ (right). The sound speed value is given by $v^2_s=\omega=1/2$. }\label{Fig5}
\end{figure*}

In this case, the fluid corresponds to an ultra-relativistic model with $\omega = 1/2$, implying the equation of state $p = \rho/2$. The critical velocity in this case is given by $	{\rm v}_c = \pm \sqrt{1/2}$. Using Eq. \eqref{rc}, the critical radius for $l > 0$ is found to be $r_c = 2.25$, which lies outside the event horizon located at $r_h = 1.8$. The Hamiltonian evaluated at the critical radius takes the value $	\mathcal{H}_c = 0.186234$ 
and the corresponding phase space portrait is depicted in Fig.\,\ref{F5a}.
From the red and blue contour lines in the phase diagram, it is evident that the dynamical system evolves from purely supersonic accretion, with velocities $-1 < {\rm v} < -{\rm v}_c$, towards subsonic flow regimes within the interval $-{\rm v}_c < {\rm v} < {\rm v}_c$. Similarly, supersonic ejection occurs for $1 > {\rm v} > {\rm v}_c$. Importantly, the fluid velocity approaches zero at the horizon, ensuring that the Hamiltonian remains constant throughout the flow.

The critical curve, represented by the lilac color in the phase diagram, illustrates the following fluid dynamics.

The lower portion of the graph depicts the evolution of the system from supersonic accretion towards the critical point at $(  2.25,  - \sqrt{1/2})$, followed by subsonic accretion continuing down to the horizon, where the fluid velocity vanishes. More specifically, the flow transitions from purely supersonic accretion through the critical point into subsonic accretion, eventually reaching the horizon with zero velocity. The system further exhibits purely supersonic accretion that passes the critical point, then transitions into subsonic accretion near the horizon at $(r_c, -{\rm v}_c)$, and finally resumes supersonic accretion down to ${\rm v} = -1$.

Conversely, the upper portion of the graph shows the fluid evolving along the lilac curve by supersonic ejection up to the critical point at $( 2.25, \sqrt{1/2})$, followed by subsonic ejection extending to the horizon, where the velocity again cancels out. The fluid undergoes purely subsonic ejection passing through the critical point, as well as purely supersonic ejection that transitions into subsonic flow near $(r_c, {\rm v}_c)$ and then accelerates back to supersonic ejection approaching ${\rm v} = 1$.

The green contour outlines solutions that evolve from supersonic flow to subsonic flow at the horizon without crossing any critical points, for either accretion or ejection. Additionally, there exist solutions beginning subsonic and ending supersonic, likewise without passing through critical points.

The dynamical system evolution for $l < 0$ is qualitatively identical to the case described above, as shown in Fig.\,\ref{F5b}, with the difference that the critical radius shifts according to the parameter $l$ to $r_c = 2.75$.

%%%%%%%%%%%%%%%%%%%%%%%%%%%%%%%%%%%%%%%%%%%%%%%%%%%%%%%%%%%%%%%
\subsection{Radiation fluid}
%%%%%%%%%%%%%%%%%%%%%%%%%%%%%%%%%%%%%%%%%%%%%%%%%%%%%%%%%%%%%%%

\begin{figure*}[ht!]
\centering
\subfigure[$l>0$ with $r_h=1.8$] 
{\label{F6a}\includegraphics[width=8.5cm]{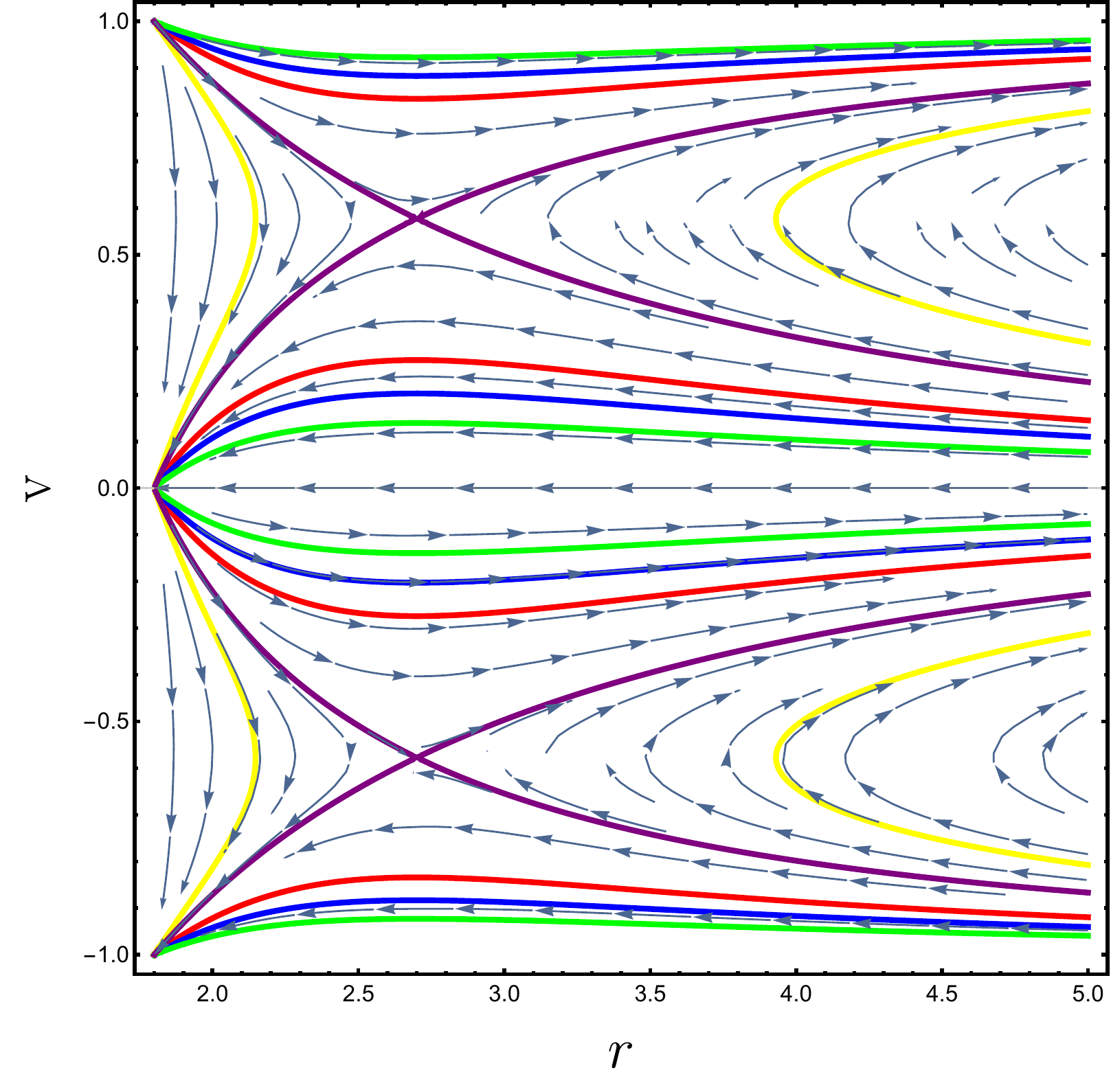} }
\hspace{0.1cm}
\subfigure[$l<0$ with $r_h=2.2$] 
{\label{F6b}\includegraphics[width=8.5cm]{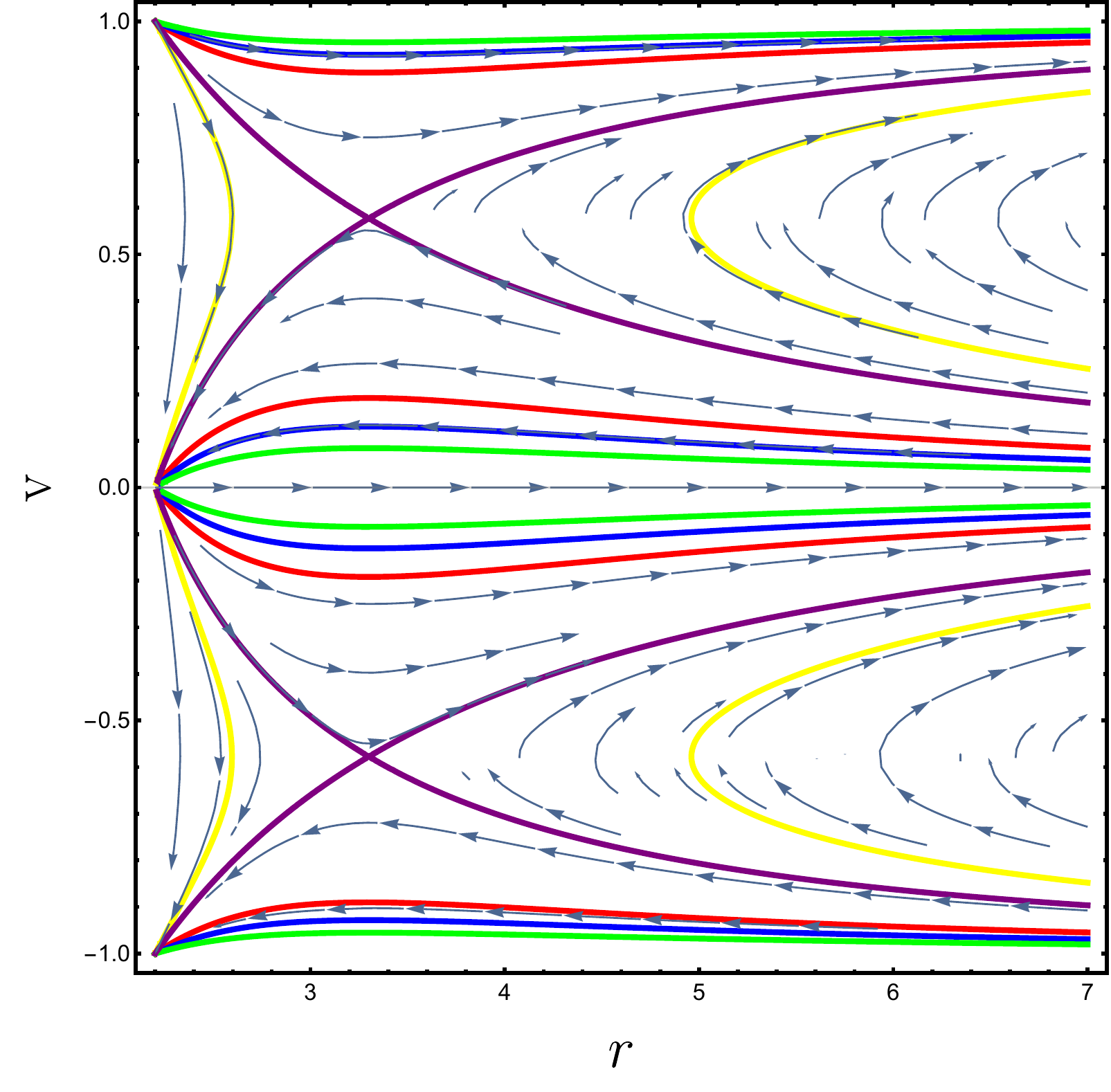} }
\hspace{0.1cm}
%\subfigure[]
%{\label{uc}\includegraphics[width=8.5cm]{M13.pdf} }
%\hspace{0.1cm}
%\subfigure[]
%{\label{ud}\includegraphics[width=8.5cm]{M14.pdf} }
%\hspace{0.1cm}
\caption{Representation of the phase space of the solution \eqref{KR0} for radiation fluid, $\omega=1/3$, and $l>0$ (left) and $l>0$ (right). The sound speed value is given by $v^2_s=\omega=1/3$.}\label{Fig6}
\end{figure*}

We now explore the behavior of the radiation fluid $(\omega=1/3)$ based on the Hamiltonian system. The critical velocity for this case is ${\rm v}_c=\pm\sqrt{1/3}$, the critical radius obtained from \eqref{rc} for $l>0$ is $r_c=2.7$ and is located after the event horizon $r_h$, which remains the same as in the previous cases. The critical Hamiltonian is $\mathcal{H}_c=0.259243$ and the phase space of the solutions is shown in Fig. \ref{F6a}. For the solutions represented in red, blue, and green, it can be seen that the dynamic system evolves with a purely supersonic ejection and accretion at $-1<{\rm v}<{\rm v}_c$ and at $1>{\rm v}>{\rm v}_c$, respectively. In addition, there is ejection followed by purely subsonic accretion in the interval $-{\rm v}_c<{\rm v}<{\rm v}_c$.

The critical curve is represented in lilac and reveals the following behavior: in regions $r<r_c$, the evolution of the system begins with supersonic ejection passing through the critical point at $(r_c, {\rm v}_c)$ and continues to subsonic ejection up to zero velocity at the horizon, where it changes to subsonic motion up to the critical point at $(r_c, -{\rm v}_c)$ and then becomes supersonic. The opposite is true for the $r>r_c$ region. There is another behavior of the dynamical system in which the motion begins as a super-subsonic ejection, passes through the critical point at $(r_c, {\rm v}_c)$ until it reaches zero velocity at the event horizon, then evolves as subsonic accretion until the critical point $(r_c, -{\rm v}_c)$, where it returns to zero velocity far from the horizon. Finally, the other possibility is described as supersonic accretion passing through the critical point $(r_c, -{\rm v}_c)$ and continuing until it reaches ${\rm v} = -1$ at the horizon, as well as a supersonic ejection that starts at ${\rm v} = 1$ and reaches the critical point $(r_c, {\rm v}_c)$, continuing supersonically away from the horizon.

The yellow curves illustrate various evolutionary paths of the system. One scenario begins with supersonic ejection, which transitions into subsonic ejection and ultimately reaches zero velocity at the event horizon. Another trajectory shows fluid motion that starts as subsonic accretion and gradually accelerates, reaching supersonic speed and attaining its maximum velocity as it approaches the horizon. These yellow curves also highlight solutions where the fluid begins in a supersonic regime and transitions to subsonic during accretion, and vice versa during ejection—both occurring without passing through the critical point. This indicates that not all physical flows require critical behavior to evolve from one regime to another. For the case where $l<0$, the overall dynamics remain qualitatively the same and are illustrated in Fig.,\ref{F6b}. However, due to the presence of the LV parameter, the location of the critical point is shifted, modifying the geometry of the phase space accordingly.

%%%%%%%%%%%%%%%%%%%%%%%%%%%%%%%%%%%%%%%%%%%%%%%%%%%%%%%%%%%%%%%
\section{ Polytropic fluid} \label{sec6}
%%%%%%%%%%%%%%%%%%%%%%%%%%%%%%%%%%%%%%%%%%%%%%%%%%%%%%%%%%%%%%%

Let us consider now the polytropic fluid equation given by
\begin{eqnarray}
	p=\mathcal{K} n^{\Gamma}\,,\label{ppoli}
\end{eqnarray}
where $\mathcal{K}$ is a constant associated with the entropy of the fluid, and $\Gamma$ is the polytropic index that characterizes the fluid's thermodynamic behavior. To analyze the dynamics and properties of such fluids, we will follow the methodology and procedures established in Refs. \cite{Ahmed, Amed2}.

The barotropic equation of state can be written in the reduced form $\rho = f(n)$, where the energy density depends solely on the number density $n$. From the thermodynamic relations in Eq. \eqref{termo}, we find that the enthalpy is given by $h = f'(n)$, where the prime denotes differentiation with respect to $n$. Furthermore, using the same relations, we obtain $p' = n h'$, which leads to  
\begin{eqnarray}
	p' = n f''\,.
\end{eqnarray}
By integrating this expression by parts, we arrive at a general relation between the pressure and the energy density,
\begin{eqnarray}
	p = n f'(n) - f(n)\,.
\end{eqnarray}
This allows us to connect the pressure function $p = g(n)$ to the energy density $\rho = f(n)$ through the equation
\begin{eqnarray}
	n f'(n) - f(n) = g(n)\,, \label{fndif}
\end{eqnarray}
which provides a bridge between barotropic and polytropic descriptions. Now, substituting the polytropic equation of state \eqref{ppoli} into Eq. \eqref{fndif}, we get
\begin{eqnarray}
	n f'(n) - f(n) = \mathcal{K} n^\Gamma\,,
\end{eqnarray}
whose solution yields the functional form of the energy density,
\begin{eqnarray}
	\rho = f(n) = m n + \frac{\mathcal{K} n^\Gamma}{\Gamma - 1}\,,
\end{eqnarray}
where $m$ is an integration constant interpreted as the baryonic mass of the fluid's particles. Consequently, the enthalpy becomes
\begin{eqnarray}
	h = f'(n) = m + \frac{\mathcal{K} \Gamma n^{\Gamma - 1}}{\Gamma - 1}\,. \label{h2}
\end{eqnarray}
Using this expression, we can now derive the speed of sound from \eqref{som2}, yielding
\begin{eqnarray}
	v_s^2 = \frac{(\Gamma - 1) \mathcal{K} \Gamma n^{\Gamma - 1}}{m (\Gamma - 1) + \mathcal{K} \Gamma n^{\Gamma - 1}} \,. \label{som3}
\end{eqnarray}
This result expresses the speed of sound as a function of the number density and the polytropic parameters, capturing the essential thermodynamic behavior of barotropic polytropic fluids.

Note that if $\Gamma > 1$, it follows that $v_s^2 < \Gamma - 1$, and consequently, the critical speed $v_c^2 < \Gamma - 1$ whenever critical points exist in the dynamical system. To proceed further, we use Eq. \eqref{c1} to express the left-hand side in terms of a reference point $(r_0, {\rm v}_0)$. This allows us to write the conserved quantity as $r_0^4n_0^2{\rm v}_0^2A(r_0)/(1-{\rm v}_0^2)$
from which we derive the relation between $n$ and $n_0$,
\begin{equation}
	\frac{n^2}{n_0^2} = \frac{r_0^4 {\rm v}_0^2 A(r_0)}{1 - {\rm v}_0^2} \; \frac{1 - {\rm v}^2}{r^4 {\rm v}^2 A(r)} = c_1^2 \frac{1 - {\rm v}^2}{r^4 {\rm v}^2 A(r)}\,, \label{nn0}
\end{equation}
where $c_1$ is a constant derived from the conserved flow. Substituting this expression into the enthalpy formula \eqref{h2}, we obtain
\begin{eqnarray}
	h = m \left[1 + \mathcal{Z} \left( \frac{1 - {\rm v}^2}{A(r) r^4 {\rm v}^2} \right)^{(\Gamma - 1)/2} \right]\,, \label{h3}
\end{eqnarray}
where the constant $\mathcal{Z}$ is defined as
\begin{eqnarray}
	\mathcal{Z} \equiv \frac{\Gamma (c_1 n_0)^{\Gamma - 1}}{m (\Gamma - 1)}\,, \label{Zposi}
\end{eqnarray}
and is strictly positive. This formulation provides a compact way to express the enthalpy $h$ as a function of the local flow velocity ${\rm v}$, radius $r$, and the background geometry encoded in $A(r)$.

Finally, we can express the Hamiltonian of the dynamical system for the polytropic fluid using Eq. \eqref{Hlt} as
\begin{eqnarray}
	\mathcal{H}(r, {\rm v}) = \frac{A(r)}{1 - {\rm v}^2} \left[ 1 + \mathcal{Z} \left( \frac{1 - {\rm v}^2}{A(r) r^4 {\rm v}^2} \right)^{\frac{\Gamma - 1}{2}} \right]^2\,, \label{Hpoli}
\end{eqnarray}
where the constant $m$ has been absorbed into a redefinition of the time coordinate $(\bar{t})$ and the Hamiltonian $(\mathcal{H})$.

%%%%%%%%%%%%%%%%%%%%%%%%%%%%%%%%%%%%%%%%%%%%%%%%%%%%%%%%%%%%%%%
\subsection{Critical points}
%%%%%%%%%%%%%%%%%%%%%%%%%%%%%%%%%%%%%%%%%%%%%%%%%%%%%%%%%%%%%%%

The critical points of the dynamical system are obtained by combining Eqs. \eqref{c1} and \eqref{Zposi}, which leads to the relation
\begin{eqnarray}
	\mathcal{K} \Gamma n^{\Gamma - 1} = m(\Gamma - 1) \mathcal{Z} \left( \frac{1 - {\rm v}^2}{A(r) r^4 {\rm v}^2} \right)^{\frac{\Gamma - 1}{2}}.
\end{eqnarray}
Using the expression for the speed of sound squared from \eqref{som3} together with the critical speed condition from \eqref{vsomcrit}, we obtain a coupled system of equations that determine the critical points $(r_c, {\rm v}_c)$:
\begin{eqnarray}
	&& {\rm v}_c^2 = \mathcal{Z} \left(\Gamma - 1 - {\rm v}_c^2 \right) \left( \frac{1 - {\rm v}_c^2}{A(r_c) r_c^4 {\rm v}_c^2} \right)^{\frac{\Gamma - 1}{2}}, \label{vcrc61}\\
	&& {\rm v}_c^2 = \frac{r_c A'(r_c)}{r_c A'(r_c) + 4 A(r_c)}. \label{vcrc62}
\end{eqnarray}
For a given value of the constant $\mathcal{Z}$, these equations can be solved simultaneously to find the location $r_c$ and velocity ${\rm v}_c$ of any existing critical points of the system. The existence and nature of these critical points depend on the fluid parameters, encoded in $\Gamma$ and $\mathcal{Z}$, as well as on the geometry through the metric function $A(r)$.

%%%%%%%%%%%%%%%%%%%%%%%%%%%%%%%%%%%%%%%%%%%%%%%%%%%%%%%%%%%%%%%
\subsection{Polytropic fluids accretion in KR black hole}
%%%%%%%%%%%%%%%%%%%%%%%%%%%%%%%%%%%%%%%%%%%%%%%%%%%%%%%%%%%%%%%

\begin{figure*}[ht!]
\centering
\subfigure[$l>0$ with $r_h=1.8$] 
{\label{P1a}\includegraphics[width=8.5cm]{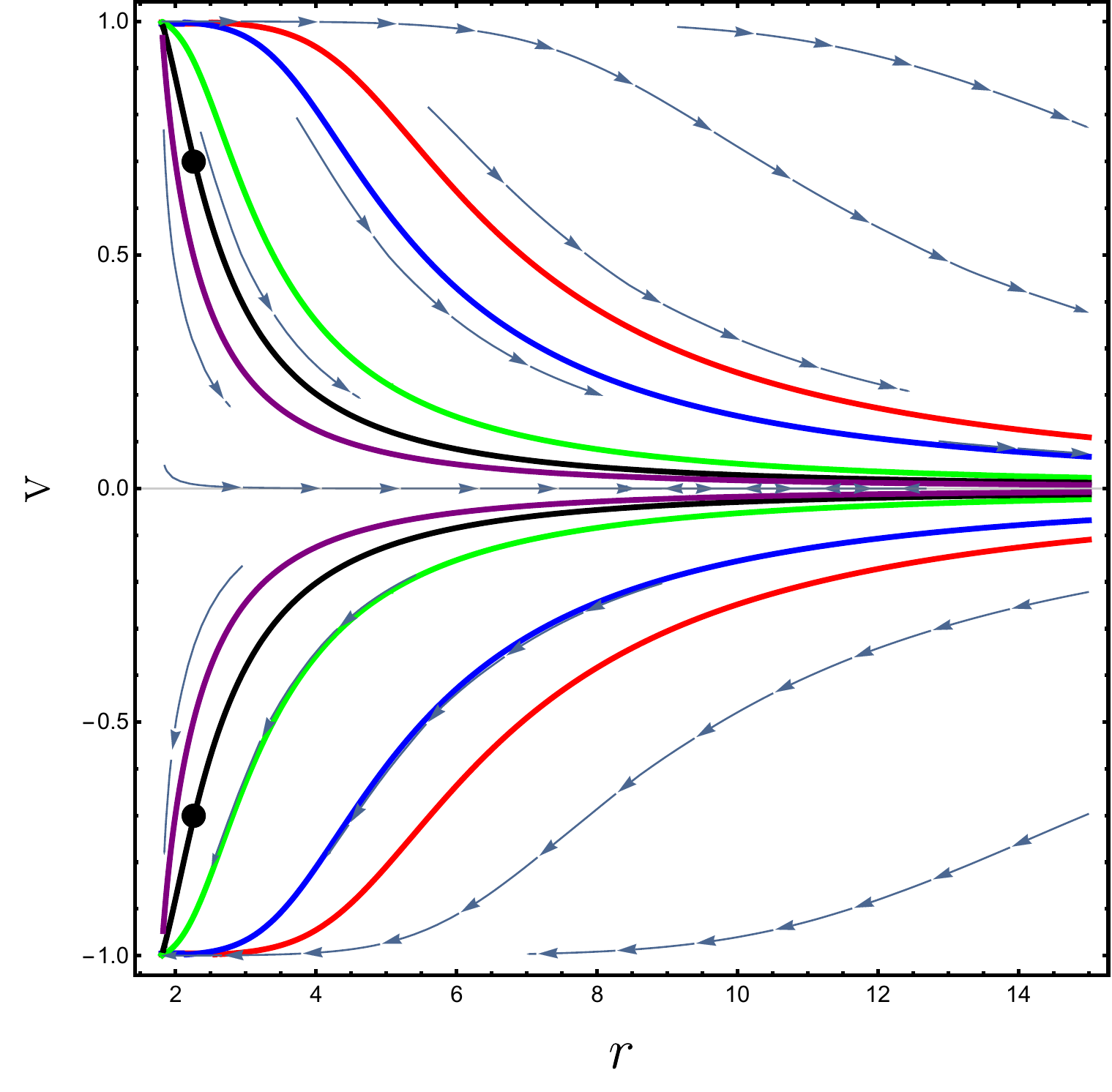} }
\hspace{0.1cm}
\subfigure[$l<0$ with $r_h=2.2$] 
{\label{P2b}\includegraphics[width=8.5cm]{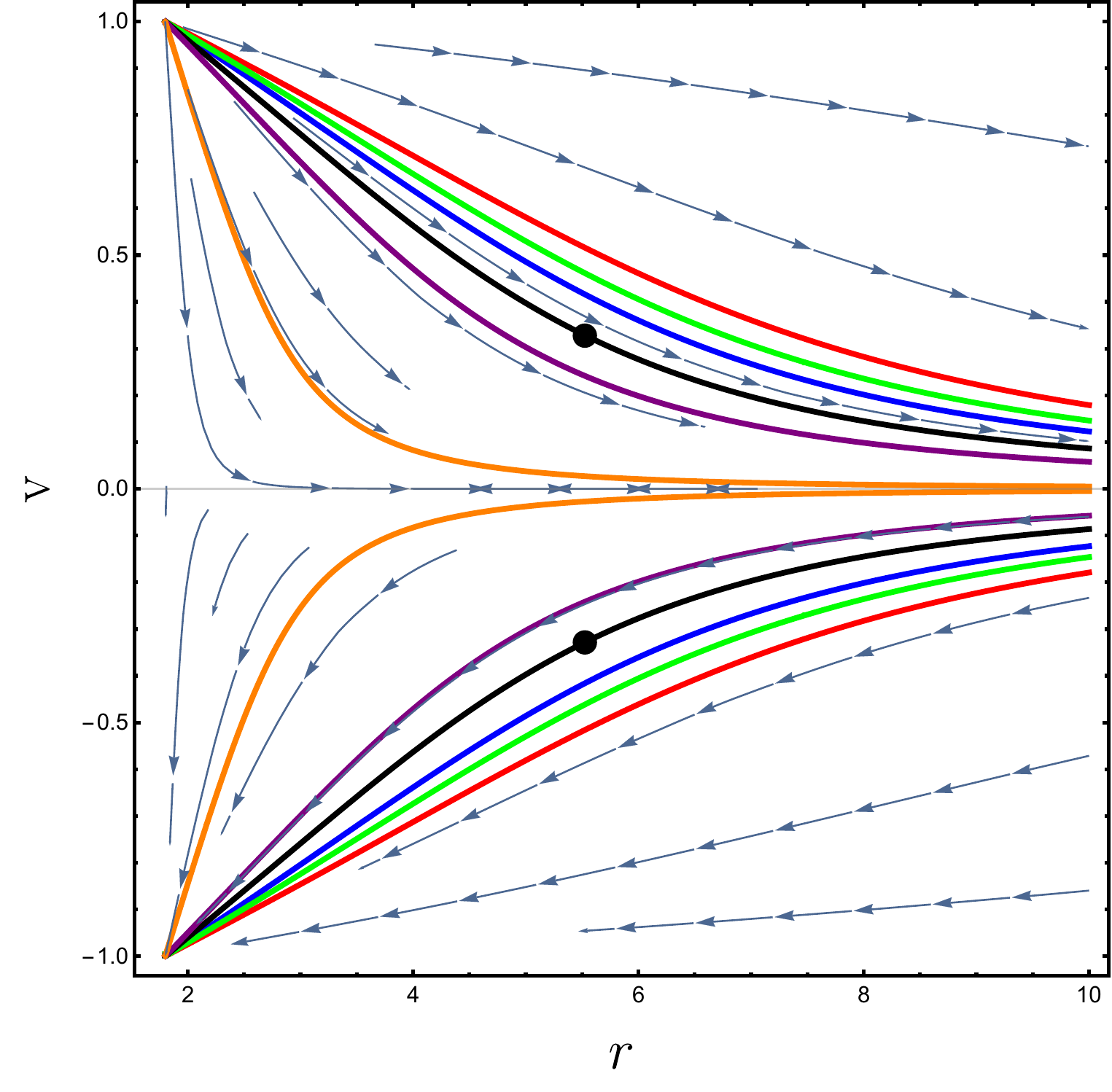} }
\hspace{0.1cm}
%\subfigure[]
%{\label{uc}\includegraphics[width=8.5cm]{M13.pdf} }
%\hspace{0.1cm}
%\subfigure[]
%{\label{ud}\includegraphics[width=8.5cm]{M14.pdf} }
%\hspace{0.1cm}
\caption{Representation of the phase space of the solution \eqref{KR0} for polytropic fluid with $\Gamma=5/3$ in \ref{P1a} and for $\Gamma=4/3$ in \ref{P2b}, and $l>0$ (left) and $l>0$ (right).
}\label{Fig7}
\end{figure*}

We will consider two types of polytropic fluids: one with $\Gamma = 5/3$, which corresponds to a non-relativistic ideal monoatomic gas, and another with $\Gamma = 4/3$, characterizing a relativistic gas. Using the metric \eqref{KR0}, we can write the Hamiltonian \eqref{Hpoli} explicitly as
\begin{eqnarray}
	\mathcal{H}(r, {\rm v}) &=& \left(\frac{1}{1 - l} - \frac{2M}{r}\right) \left(1 - {\rm v}^2\right)^{-1} \nonumber \\
	&& \times \left[ 1 + \mathcal{Z} \left( \frac{r^4 {\rm v}^2 (1 - {\rm v}^2)}{\frac{1}{1 - l} - \frac{2M}{r}} \right)^{\frac{1}{2} (\Gamma - 1)} \right]^2. \label{Hpoli2}
\end{eqnarray}
The evolution of the dynamical system describing the polytropic fluid with $\Gamma = 5/3$, in the presence of the LV parameter of the KR field $l=0.1$, is illustrated by the trajectories plotted in Fig.\,\ref{P1a}.

The critical points of the system are indicated by black dots at $(r_c, {\rm v}_c) = (2.26,\,0.70)$, through which the black critical curve passes. The fluid dynamics evolve as follows: in the upper region of the phase space, the fluid begins with a supersonic ejection, crosses the critical point, and then slows down to a subsonic ejection far from the horizon, which is located at $r_h = 1.8$, where the velocity reaches zero. It is worth noting that the critical point lies quite close to the horizon. 

In the lower part of the graph, the fluid starts its accretion flow at a subsonic speed far from the horizon and gradually accelerates to supersonic speeds, reaching its maximum velocity as it approaches the horizon. The figure illustrates the case with the LV parameter $l > 0$. For $l < 0$, the dynamical behavior remains essentially the same, with the only difference being a shift of the horizon position to $r_h = 2.2$.

For the relativistic fluid, characterized by $\Gamma=4/3$, the evolution of the dynamical system is depicted by the curves shown in Fig.\,\ref{P2b}, with the LV parameter set to $l=0.1$. The behavior observed here closely resembles that of the previous case: the system transitions from supersonic to subsonic ejection far from the event horizon, followed by subsonic to supersonic accretion as the fluid approaches the horizon, ultimately reaching its maximum velocity there. 
In this scenario, the critical points are located at $(r_c, {\rm v}_c) = (5.52,\,0.32)$, which lie considerably farther from the event horizon compared to the non-relativistic case. The unique critical solution corresponding to these points is illustrated by the black curve in the figure.

Therefore, in the first solution, the fluid flow transitions to supersonic speeds very close to the black hole horizon. Before reaching this point, the fluid remains in a subsonic regime, which indicates that the flow is strongly influenced by the curvature of spacetime near the black hole. This close proximity to the horizon means that gravitational effects dominate the fluid dynamics in this region. In contrast, in the second case, the transition from subsonic to supersonic flow occurs much farther away from the black hole. As a result, the fluid attains high velocities at a relatively large distance from the black hole, where spacetime curvature effects are less pronounced. The location of the critical point, therefore, plays a significant role in determining the efficiency of accretion, as it affects how and where the fluid accelerates and interacts with the gravitational field of the black hole.

%%%%%%%%%%%%%%%%%%%%%%%%%%%%%%%%%%%%%%%%%%%%%%%%%%%%%%%%%%%%%%%
\section{Summary and  conclusion}\label{Sec:Conclusion}
%%%%%%%%%%%%%%%%%%%%%%%%%%%%%%%%%%%%%%%%%%%%%%%%%%%%%%%%%%%%%%%
 
In this work, we have considered a spherically symmetric and static Schwarzschild-like metric solution that incorporates the Kalb-Ramond (KR) field, modified by a spontaneous Lorentz symmetry breaking parameter \cite{KR}. We investigated spherically symmetric accretion onto the KR black hole for three types of isothermal fluids governed by the equation of state $p=\omega \rho$, characterized by the state parameter $\omega$, namely ultra-stiff fluid $(\omega=1)$, ultra-relativistic fluid $(\omega=1/2)$, and radiation $(\omega=1/3)$.

The influence of the LV parameter $l$ on the fluid density $\rho(r)$, radial velocity $u(r)$, and accretion rate $\dot{M}$ was analyzed in Sec.\,\ref{sec3}, following the description of the metric and relevant accretion equations in Sec.\,\ref{sec1}. The results presented in Fig.\,\ref{Fig1} showed that for positive (negative) values of $l$, the fluid density increases (decreases) as compared to the Schwarzschild case; in both scenarios, the density approaches zero at spatial infinity. Fig. \ref{Fig2} demonstrated that for all fluid types considered, the radial velocity exhibits a transition from subsonic to supersonic regimes. Moreover, the LV parameter $l$ affects the location of this transition: when $l$ is negative, the transition occurs farther from the black hole relative to the Schwarzschild solution, while for positive $l$ it moves closer to the event horizon.
Regarding the accretion rate, the parameter $l$ was found to enhance (suppress) the rate for positive (negative) values. Nevertheless, in all cases examined, the total mass of the black hole increased as a consequence of matter accretion.

To evaluate the behavior of the fluid near and at the critical points, we employed the dynamical system approach through the Hamiltonian formalism, as adopted in several previous works \cite{Ahmed, ednaldo1, Amed2, Abdul}. To this end, we first revisited this formalism in Sec. \ref{Sec3}. For the fluid with $\omega=1$, our result was expected: we observed that the dynamical system exhibits no critical points. The phase space diagram of the solutions is shown in Fig.\,\ref{Fig4}, where purely subsonic accretion $({\rm v}<0)$ or purely subsonic ejection $({\rm v}>0)$ is observed. The physical flow occurs for three-velocity magnitudes less than one, confined between the lilac curves where the Hamiltonian attains its minimum value. The influence of the parameter $l$ in this case manifests in the radius of the horizon, which also serves as the critical radius, increasing or decreasing depending on the sign of $l$. For ultra-relativistic fluids with $l>0$, the solution curves are depicted in Fig.\,\ref{Fig5}. The critical points are located at $(2.25, \pm\sqrt{1/2})$, positioned outside the event horizon, and correspond to the lilac critical curve. In this case, in the lower portion of the graph, the fluid begins with supersonic accretion that passes through the critical point, transitions to subsonic flow, and continues until the horizon where the fluid velocity vanishes. Additionally, there exists a solution that evolves from purely supersonic to subsonic accretion, reaching the maximum speed at the horizon.

In the upper part of the graph, we observed a supersonic ejection that passed through the critical point, after which the system evolved to subsonic flow until the horizon where the fluid velocity vanishes. There was also a purely subsonic ejection followed by a purely supersonic ejection. No differences in fluid behavior were found for $l<0$; it was identical to the case described here. However, the critical radius was shifted, indicating that the transition through the sonic points occurred further away from the black hole. For the radiation fluid, the evolution of the dynamical system is shown in Fig.\,\ref{Fig6}. The critical curve, shown in light color, passes through the critical points $(2.7,\,\pm\sqrt{1/3})$ for the case with $l>0$ and was located beyond the event horizon. At the top of the graph, a supersonic ejection passes through the sonic points and continues as subsonic ejection until reaching zero velocity at the horizon. At the bottom of the graph, subsonic accretion proceeded up to the sonic point and then transitioned to supersonic accretion. There was also a sonic ejection passing through the sonic point to zero velocity at the horizon, then evolving as subsonic accretion back to the sonic point, where the velocity returns to zero far from the horizon. Additionally, supersonic accretion passes through the sonic point and continues until reaching maximum speed at the horizon, while supersonic ejection begins at maximum speed, passes through the sonic point, and then becomes supersonic far from the horizon.

In Sec.\,\ref{sec6}, we described the evolution of non-relativistic ideal monoatomic polytropic fluids with $\Gamma=5/3$ and relativistic gases with $\Gamma=4/3$, starting from the dynamical system. The Hamiltonians for these types of fluids were re-derived. Our results showed that, in both cases considered, there is a black curve solution, plotted in Fig.\,\ref{Fig7}, which passes through the critical points of the system, marked by black dots on the graphs and located beyond the event horizon. In both scenarios, in the upper part of the graph, the system evolves from supersonic ejection through the sonic point to subsonic ejection, where the velocity vanishes far from the horizon; in the lower part of the graph, the fluid transitions from subsonic accretion through the critical point to supersonic accretion until it reaches the speed of light at the horizon. The effect of the LV parameter in this case is to shift the horizons closer to the origin for $l>0$, or further away from it for $l<0$. Therefore, the solutions presented in Fig.\,\ref{Fig7} exhibited the same qualitative behavior for both positive and negative values of $l$. It was also noted that, for the non-relativistic case, the fluid became supersonic very close to the event horizon, thus experiencing a stronger influence of spacetime curvature. In contrast, for the relativistic fluid, the transition to supersonic flow occurred further from the event horizon. Our results demonstrate that the accretion of matter by black holes can be influenced by the Lorentz symmetry breaking parameter when it is present in the solution.

Finally, we analyzed the spherical accretion of various fluid types. By applying a dynamical systems approach and Hamiltonian formalism, we characterized fluid behavior for both isothermal and polytropic equations of state, identifying critical points and flow regimes influenced by the symmetry breaking parameter. Our findings revealed that this parameter affects the transitions between subsonic and supersonic flow, and the overall accretion efficiency, with notable differences observed between non-relativistic and relativistic fluids. 

To conclude, the results results obtained in this work highlight the significant role that Lorentz symmetry breaking can play in black hole accretion dynamics. For future work, it would be of great interest to extend this analysis to rotating black hole spacetimes, where frame dragging is expected to introduce new effects on accretion flows. Additionally, studying non-spherical, axisymmetric accretion models, or including magnetic fields could provide a more realistic description of astrophysical accretion processes in Lorentz-violating gravity theories. Such investigations could deepen our understanding of observational signatures and constraints on Lorentz symmetry breaking in strong gravity regimes.

%%%%%%%%%%%%%%%%%%%%%%%%%%%%%%%%%%%%%%%%%%%%%%%%%%%%%%%%%%%%%%%
\section*{Acknowledgements}
%%%%%%%%%%%%%%%%%%%%%%%%%%%%%%%%%%%%%%%%%%%%%%%%%%%%%%%%%%%%%%%

MER thanks Conselho Nacional de Desenvolvimento Cient\'ifico e Tecnol\'ogico - CNPq, Brazil, for partial financial support. This study was financed in part by the Coordena\c{c}\~{a}o de Aperfei\c{c}oamento de Pessoal de N\'{i}vel Superior - Brasil (CAPES) - Finance Code 001.
FSNL acknowledges support from the Funda\c{c}\~{a}o para a Ci\^{e}ncia e a Tecnologia (FCT) Scientific Employment Stimulus contract with reference CEECINST/00032/2018, and funding through the research grants UIDB/04434/2020, UIDP/04434/2020 and PTDC/FIS-AST/0054/2021.
DRG is supported by the Agencia Estatal de Investigación Grant Nos. PID2022-138607NB-I00 and CNS2024-154444, funded by MICIU/AEI/10.13039/501100011033 (Spain).

%%%%%%%%%%%%%%%%%%%%%%%%%%%%%%%%%%%%%%%%%%%%%%%%%%%

%%%%%%%%%%%%%%%%%%%%%%%%%%%%%%%%%%%%%%%%%%%%%%%%%%%

%%%%%%%%%%%%%%%%%%%%%%%%%%%%%%%%%%%%%%%%%%%%%%%%%%%
\end{document}